\documentclass[aps,prb,twocolumn,english,showpacs,superscriptaddress,amssymb,amsfonts]{revtex4}
\usepackage[T1]{fontenc}
\usepackage[latin9]{inputenc}
\usepackage{amsmath}
\usepackage{dsfont}
\usepackage{amstext}

\usepackage{tocvsec2}

\usepackage{amssymb}
\usepackage{amsbsy}
\usepackage{amsthm}
\usepackage{epsfig}
\usepackage{framed}
\usepackage{graphicx}
\usepackage{bbm}
\usepackage{hyperref}
\usepackage{color}
\usepackage{multirow}

\newcommand{\bst}{{\mathcal{T}}}

\newcommand{\ie}{{\emph{i.e.~}}}
\makeatletter

\newcommand{\Rmnum}[1]{\expandafter\@slowromancap\romannumeral #1@}
\makeatother
\newcommand{\imth}{\hspace{1pt}\mathrm{i}\hspace{1pt}}

\newcommand{\mbz}{{\mathbb{Z}}}
\newcommand{\bea}{\begin{eqnarray}}
\newcommand{\eea}{\end{eqnarray}}
\newcommand{\bpm}{\begin{pmatrix}}
\newcommand{\epm}{\end{pmatrix}}
\newcommand{\bal}{\begin{aligned}}
\newcommand{\eal}{\end{aligned}}

\newcommand{\dket}[1]{|{#1}\rangle}

\makeatother

\usepackage{babel}
\begin{document}
\title{Field induced quantum spin liquid with spinon Fermi surfaces in the Kitaev model}

\author{Hong-Chen Jiang}
\affiliation{Stanford Institute for Materials and Energy Sciences, SLAC and Stanford University, Menlo Park, California 94025, USA}
\author{Chang-Yan Wang}
\affiliation{Department of Physics, The Ohio State University, Columbus, OH 43210, USA}
\author{Biao Huang}
\affiliation{Department of Physics and Astronomy, University of Pittsburgh, Pittsburgh PA 15260, USA}
\author{Yuan-Ming Lu}
\affiliation{Department of Physics, The Ohio State University, Columbus, OH 43210, USA}

\begin{abstract}
Recent experimental evidence for a field-induced quantum spin liquid (QSL) in $\alpha$-RuCl$_3$ calls for an understanding for the ground state of honeycomb Kitaev model under a magnetic field. In this work we address the nature of an enigmatic gapless paramagnetic phase in the antiferromagnetic Kitave model, under an intermediate magnetic field perpendicular to the plane. Combining theoretical and numerical efforts, we identify this gapless phase as a $U(1)$ QSL with spinon Fermi surfaces. We also reveal the nature of continuous quantum phase transitions involving this $U(1)$ QSL, and obtain a phase diagram of the Kitaev model as a function of bond anisotropy and perpendicular magnetic field.
\end{abstract}


\maketitle
%\setcounter{tocdepth}{2}
%\begin{widetext}
%\tableofcontents
%\end{widetext}

%\maxsecnumdepth{subsection}

%{\small \setcounter{tocdepth}{2} \tableofcontents}

%\twocolumn

\section{Introduction}

The quest for quantum spin liquids (QSLs) in frustrated magnetic materials has been a longstanding challenge in modern condensed matter physics\cite{Balents2010,Lee2014a,Savary2017,Zhou2017}. The exact QSL ground state in Kitaev's solvable bilinear spin-$1/2$ honeycomb model\cite{Kitaev2006} leads to the possibility of realizing QSLs in a large family of layered Mott insulators with strong spin-orbit couplings, coined ``Kitaev materials''\cite{Jackeli2009,Plumb2014,Kim2015a,Witczak-Krempa2014,Rau2016,Winter2017,Trebst2017,Hermanns2018,Jang2018}. Among them, $\alpha$-RuCl$_3$ is a promising Kitaev material consisting of effective spin-$1/2$s on distorted honeycomb layers. Although the material exhibits a ``zigzag'' magnetic order below $T_N=7\sim14$ K, recent experimental efforts showed that applying an external magnetic field can suppress the order and drive $\alpha$-RuCl$_3$ into a paramagnetic phase\cite{Yadav2016,Baek2017,Wolter2017,Leahy2017,Hentrich2018,Kasahara2018}, a plausible candidate for QSLs.

Since Kitaev-type interaction plays an important role in the effective spin model of $\alpha$-RuCl$_3$, these experimental progresses provided a strong motivation to understand the properties of honeycomb Kitaev model under a magnetic (or Zeeman) field. Indeed there has been quite some numerical efforts to study the ground states of Kitaev model under an external magnetic field\cite{Jiang2011,Zhu2018,Ronquillo2018,Hickey2018,Gohlke2018,Nasu2018,Liang2018,Lampen-Kelley2018,Lampen-Kelley2018a}. In particular when a perpendicular field $h_{[111]}$ is applied to the antiferromagnetic (AF) Kitaev model, between the non-Abelian Ising topological order at low field\cite{Kitaev2006} and fully polarized state at high field, there is an intermediate paramagnetic phase\cite{Zhu2018,Ronquillo2018,Hickey2018,Gohlke2018} which appears to be gapless. What is the nature of this field-induced enigmatic gapless state?

The goal of this paper is to address this question, and to understand the nature of quantum phase transitions in the AF Kitaev model under a [111] magnetic field. By combining symmetry analysis, topological classification, analytical perturbation theory and numerical studies, we identify a symmetric $U(1)$ spin liquid with spinon Fermi surfaces (FSs), termed $U1A_{k=0}$ state, as the only candidate state for the gapless phase under the intermediate field. In a two-dimensional phase diagram (FIG. \ref{fig:schemaic pd}) for Kitaev model as a function of bond anisotropy $J_z/J_{x,y}$ and perpendicular field $h_{[111]}$, we unify 4 quantum phases including Abelian toric code phase, non-Abelian Ising phase, gapless $U(1)$ QSL and the trivial polarized phase. We also provide the low-energy effective theories describing the continuous phase transitions between these 4 phases.

\section{The model and its phase diagram}

We study the following anisotropic Kitaev model with AF couplings $J_\alpha>0$, under a perpendicular magnetic field along $[111]$ direction
\bea\label{model}
\hat H_{K}(\vec h)=\sum_{\langle i,j\rangle}J_{\alpha_{ij}}S^{\alpha_{ij}}_iS^{\alpha_{ij}}_j-\vec h\cdot\sum_{i}{\bf S}_i,\\
\notag J_x=J_y=J,~~~\vec h={h}(1,1,1).
\eea
where $\alpha_{ij}=x,y,z$ for the nearest neighbors (NNs) $\langle i,j\rangle$ along 3 different orientations.

%\begin{figure}
%\includegraphics[width=0.6\columnwidth]{lattice}
%\caption{(color online) The unit cell and crystal symmetries of the honeycomb Kitaev model (\ref{model}).}
%\label{fig:lattice}
%\end{figure}

\begin{figure}
\includegraphics[width=\columnwidth]{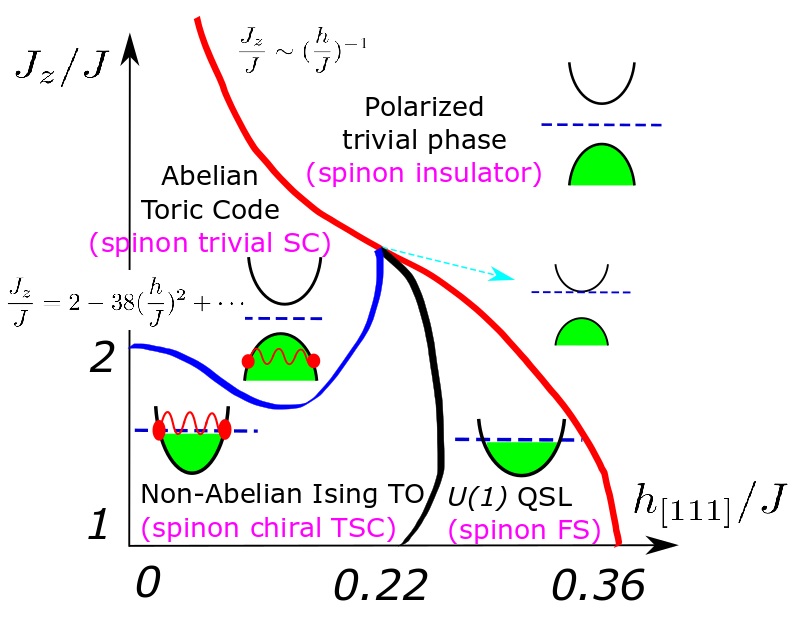}
\caption{(color online) Schematic phase diagram of antiferromagnetic Kitaev model (\ref{model}) with anisotropy under a [111] magnetic field, and the ``spinon fermiology'' of all 4 phases therein. In particular, there is a quadrucritical point separating the 4 phases, where the three phase boundaries (labeled by red, blue and green colors) intersect.}
\label{fig:schemaic pd}
\end{figure}

\begin{figure}[h]
\centering
(a)\includegraphics[width=0.4\columnwidth]{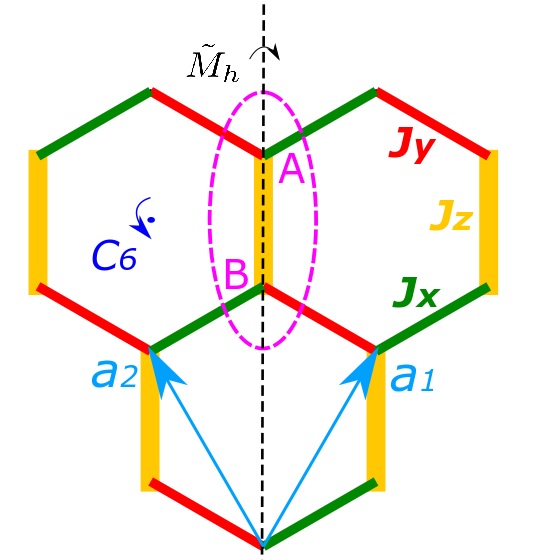}(b)\includegraphics[width=0.53\columnwidth]{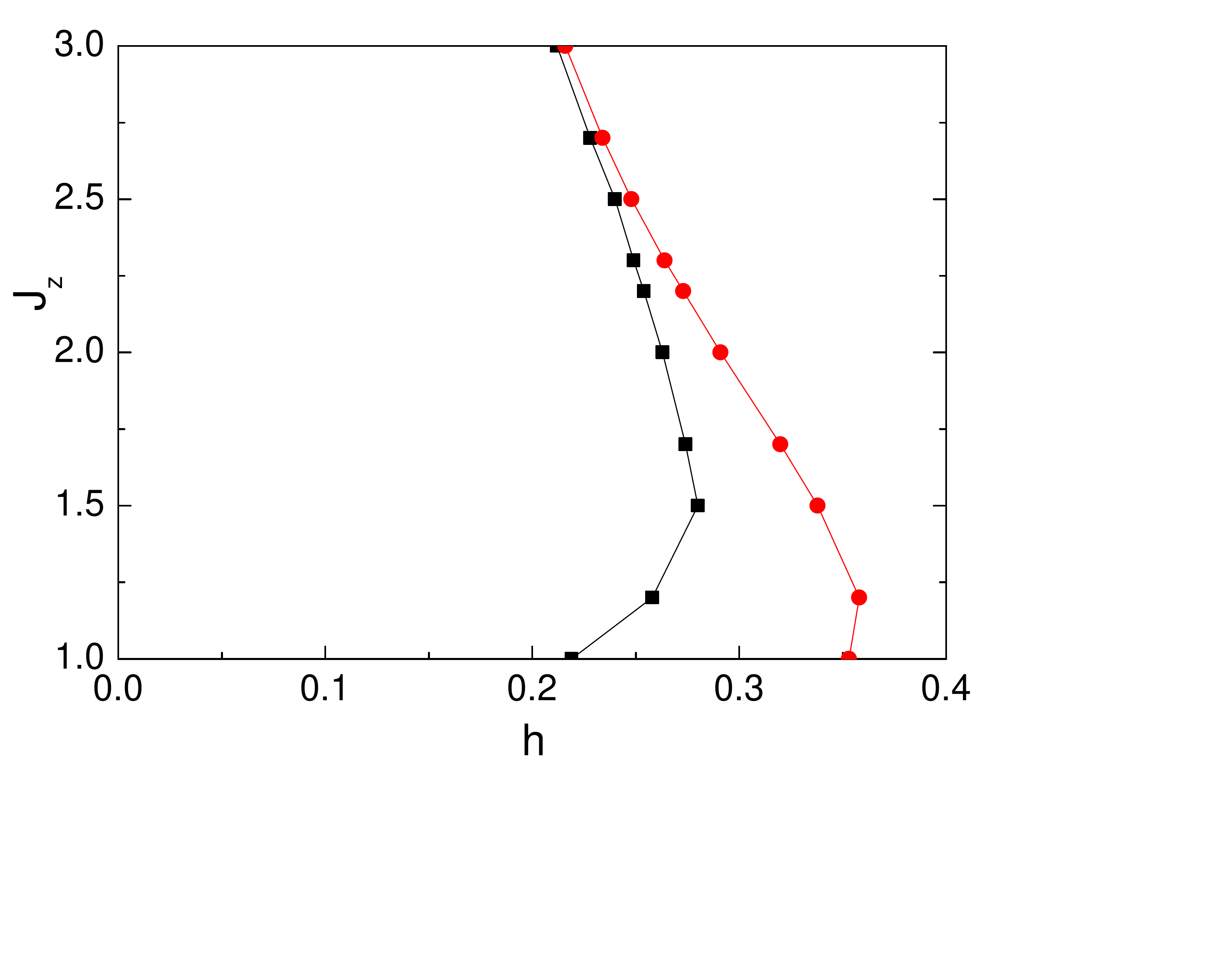}
\caption{(Color online) (a) The unit cell and crystal symmetries of the honeycomb Kitaev model (\ref{model}). (b) Phase boundaries determined by ED calculations on a $N=24$ torus with $L_x=4$ and $L_y=3$.}
\label{fig:lattice+pd}
\end{figure}

Under a uniform magnetic field along $[111]$ direction, the isotropic Kitaev model ($J_\alpha=J$) preserves two translational symmetries $T_{1,2}$ and 6-fold rotational symmetry $C_6$ around each hexagon center. Although time reversal $\bst$ is broken by the magnetic field, the combination $\tilde M_h$ of time reversal and mirror reflection w.r.t. $[1\bar10]$ plane is still preserved (see FIG. \ref{fig:lattice+pd}(a))
\bea
\tilde M_h=M_{[1\bar10]}\cdot\bst
\eea
Labeling each spin $(x_1,x_2,s)$ by its Bravais lattice vector ${\bf r}=x_1\vec a_1+x_2\vec a_2$ and sublattice index $s=0,1$ (for A/B sublattices), it transforms under the two point group symmetries as follows
\bea
&(S^x,S^y,S^z)_{(x_1,x_2,s)}\overset{\tilde M_h}\longrightarrow(S^y,S^x,S^z)_{(x_2,x_1,s)},\\
&(S^x,S^y,S^z)_{(x_1,x_2,s)}\overset{C_6}\longrightarrow(S^z,S^x,S^y)_{(1-x_2,x_1+x_2-s,1-s)}.~~~
\eea
The anisotropy $J_z\neq J_{x,y}=J$ breaks the 6-fold rotation $C_6$ but preserves the inversion symmetry $I=(C_6)^3$ w.r.t to the hexagon center:
\bea\label{sym:inversion}
{\bf S}_{(x_1,x_2,s)}\overset{I}\longrightarrow{\bf S}_{(1-x_1,1-x_2,1-s)}
\eea
As will become clear later, these symmetries play an important role in determining the phase diagram (FIG. \ref{fig:schemaic pd}).

As shown by Kitaev\cite{Kitaev2006}, at small field $h\ll J_\alpha$ the model is exactly solvable in the Majorana representation of spin-$1/2$ operators:
\bea\label{maj rep}
2\hat S^\alpha=\imth b^\alpha c=-\imth\frac{\epsilon_{\alpha\beta\gamma}}2b^\beta b^\gamma.
\eea
where 4 Majorana fermions $\{b_i^\alpha,c_i|\alpha=x,y,z\}$ are introduce for every site. In particular, the small-field ground state of model (\ref{model}) can be obtained by enforcing the following constraint for each site
\bea\label{maj rep:constraint}
b^x_ib_i^yb_i^zc_i=1,~~~\forall~i.
\eea
on the ground state of mean-field Hamiltonian:
\bea
\notag\hat H_{MF}^{Z_2}=-\sum_{\langle i\in A,j\in B\rangle}\big(\frac{\Delta_{2,\alpha_{ij}}}2\imth b_i^{\alpha_{ij}}b_j^{\alpha_{ij}}+\frac{\tilde J_{\alpha_{ij}}}4\imth c_ic_j\big)\\
\label{maj rep:mfh}-{\imth h}\sum_i(b_i^x+b_i^y+b_i^z)c_i+\frac{g_3h^3}{8J^2}\sum_{\langle\langle i,j\rangle\rangle}\nu_{ij}\imth c_ic_j+O(\frac{h^3}{J^2})~
\eea
where $\langle\langle i,j\rangle\rangle$ denotes a pair of next nearest neighbors (NNNs) and $\nu_{ij}=\pm1$ labels the clockwise hopping sign (around each hexagon center) between two NNNs.

Majorana representation (\ref{maj rep}) has a one-to-one correspondence with the more familiar Abrikosov-fermion representation\cite{Abrikosov1965,Wen2002,Burnell2011} of spin-$1/2$'s:
\bea
\notag{\bf S}_i=\frac14\text{Tr}\big(\Psi^\dagger_i\vec{\sigma}\Psi_i\big),~~f_{i\uparrow}=\frac{b_i^z+\imth c_i}2,~~f_{i\downarrow}=\frac{b_i^x+\imth b_i^y}2,\\
\Psi_i=\bpm f_{i\uparrow}&f^\dagger_{i\downarrow}\\f_{i\downarrow}&-f_{i\uparrow}^\dagger\epm=\frac12\big(\sum_{\alpha=x,y,z}b_i^\alpha\hat\sigma_\alpha+\imth c~\hat 1_{2\times2}\big).~~~\label{af rep}
\eea
where constraint (\ref{maj rep:constraint}) becomes the single-occupancy constraint for fermionic spinons $\{f_{i,\uparrow/\downarrow}\}$:
\bea\label{af rep:constraint}
f^\dagger_{i\uparrow}f_{i\uparrow}+f^\dagger_{i\downarrow}f_{i\downarrow}=1.
\eea

This representation provides an intuitive picture to understand the phase diagram of model (\ref{model}). As shown in FIG. \ref{fig:schemaic pd}, all 4 phases in the phase diagram can be conveniently understood by their fermiology of spinons:

(1) Gapped non-Abelian topological order (TO) of the Ising type\cite{Nayak2008}, which corresponds to a $p_x+\imth p_y$ (``weak pairing''\cite{Read2000}) chiral topological superconductor (TSC) of femrionic spinons. It is stablized by a small magnetic field along $[111]$ direction, when the anisotropy is weak \ie $J_z\simeq J_x=J_y=J$.

(2) Gapped Abelian TO of the toric code type\cite{Kitaev2003}, which corresponds to a trivial (``strong pairing''\cite{Read2000}) superconductor of fermionic spinons. It is stablized by a small magnetic field and large anisotropy. At small field $h\ll J$, the phase boundary between Abelian toric code and non-Abelian Ising phases can be analytically determined from perturbation theory (see Appendix \ref{app:ising<->toric code} for details):
\bea\label{phase bdy:ising<->toric code}
\frac{J_z}{J}\approx2-38(\frac{h}{J})^2+O(\frac hJ)^4.
\eea

(3) Gapless $U(1)$ spin liquid, which is a spinon metal with both electron and hole fermi surfaces (FS) coupled to an emergent $U(1)$ gauge field. This phase is stablized by an intermediate magnetic field and weak anisotropy. Due to the single-occupancy constraint (\ref{af rep:constraint}) the spinons have an integer filling number (2 per unit cell), and as a result the spinon FS (see FIG. \ref{fig:spinon fs}) at isotropy point ($J_\alpha=J$) consists of one electron-type pocket at BZ center $\Gamma$ and two hole-type pockets at BZ corners $\pm K$. Increasing magnetic field $h_{[111]}$ will shrink the size of all pockets. Meanwhile increasing bond anisotropy $J_z/J$ will also move the two hole pockets towards hexagonal BZ edge center $M$ with $k_1=k_2=\pi$, in addition to shrinking them, as illustrated in FIG. \ref{fig:spinon fs}.

(4) Gapped polarized phase, which is a trivial band insulator of spinons. In contrast to all other phases hosting fractionalized spinon excitations, here the spinons are confined due to proliferation of $U(1)$ monopoles. This phase is adiabatically connected to the trivial product state where all spins align along $[111]$ direction at a high field. In particular in the limit of small field and strong anisotropy $h/J\ll\sqrt{J/J_z}\ll1$, perturbation theory reveals the low-energy physics of model (\ref{model}) as the toric code under a transverse field\cite{Vidal2009,Vidal2009a,Dusuel2011}. Therefore the phase boundary between the polarized phase and toric code phase can be determined via perturbation theory (for details see Appendix \ref{app:polarized<->toric code}):
\bea\label{phase bdy:toric code<->polarized}
J_z/J\sim(h/J)^{-1}
\eea

\begin{figure}
  \includegraphics[width=\linewidth]{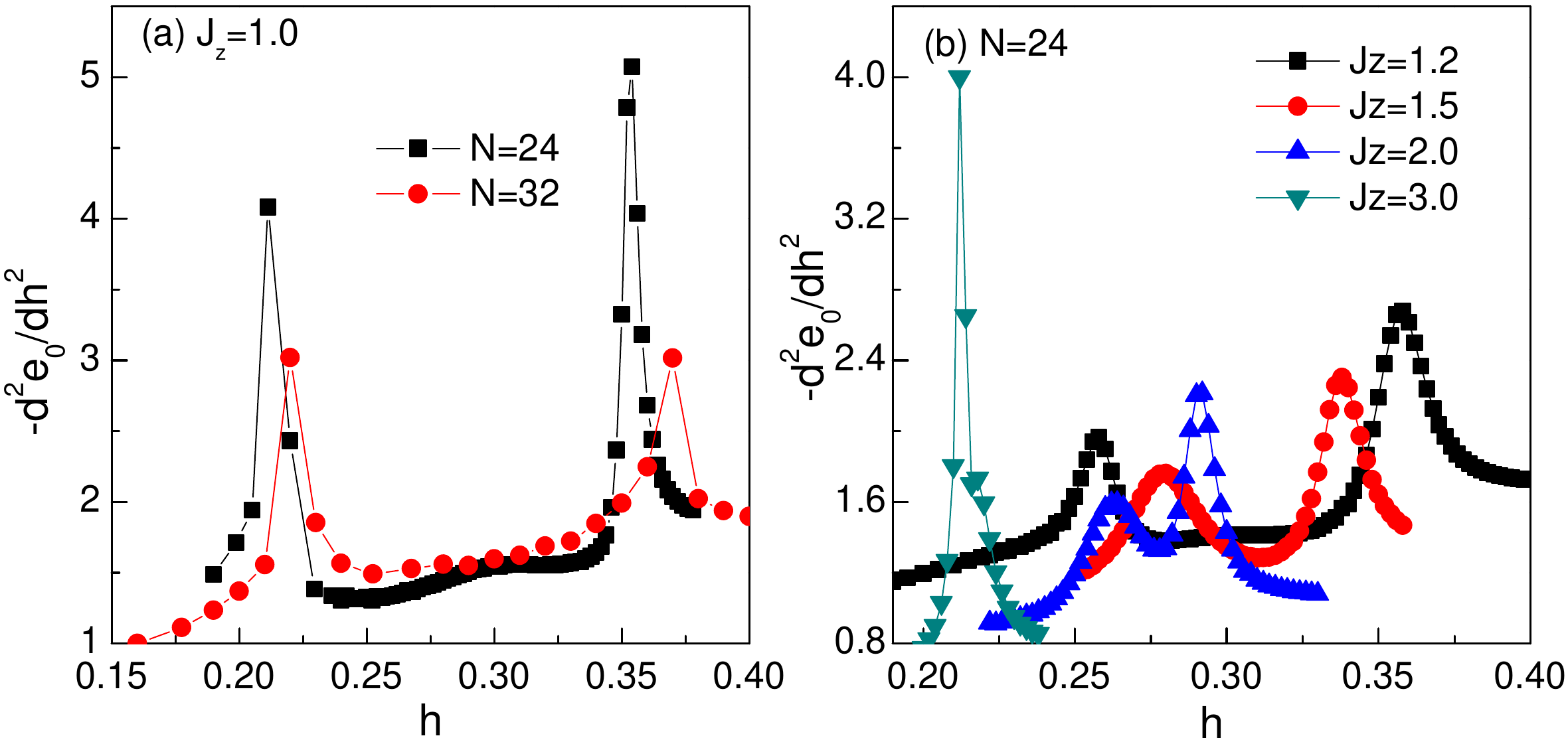}
  \caption{(Color online) Second derivative of the ground state energy density -$d^2e_0/dh^2$ of model (\ref{model}) as a function magnetic field $h$ for (a) $J_z=1.0$ on $N=24$ and $N=32$ torus, and (b) different $J_z$ on $N=24$ torus. We set $J=1$ in model (\ref{model}) for all calculations.}\label{Fig:dE2n}
\end{figure}

This schematic phase diagram is further confirmed by numerical simulations using the exact diagonalization (ED) and  and density-matrix renormalization group (DMRG)\cite{White1992} method. We consider $L_x{\bf e}_x\times L_y{\bf e}_y$ torus geometry with periodic boundary condition along both ${\bf e}_y=\vec a_1$ and ${\bf e}_x=\vec a_1-\vec a_2$ directions, with a total number of $N=2L_xL_y$ sites. ED calculations are performed on a $N=24$ torus with $L_x=4$ and $L_y=3$. DMRG calculations are performed on a $N=32$ torus with $L_x=4$ and $L_y=4$ where we keep up to $m=3072$ block states with a truncation error $\epsilon\sim 10^{-7}$. In FIG. \ref{fig:lattice+pd}(b), the phase boundaries between different phases are determined by calculating the ground state energy using ED on $L_x=4,L_y=3$ torus. As shown in Fig.\ref{Fig:dE2n}, the second derivative of the ground state energy density $-d^2e_0/dh^2$ shows two visible peaks as a function of $h$, which gives us the two phase boundaries with critical magnetic field $h_{c1}$ (black squares) and $h_{c2}$ (red circles). Compared to FIG. \ref{fig:schemaic pd}, the $h_{c1}$ separates non-Abelian Ising TO and $U(1)$ QSL (black), while $h_{c2}$ separates the $U(1)$ QSL and the polarized trivial phase. When the anisotropy becomes large enough $J_z/J\geq3$, the two phase boundaries \ie two peaks in FIG. \ref{Fig:dE2n}(b) merge into a single one as demonstrated in FIG. \ref{fig:schemaic pd}. Calculations on $L_x=L_y=4$ torus lead to the same results qualitatively, as shown in FIG. \ref{Fig:dE2n}(a).

The ground state energy of ED calculations however fails to distinguish the Abelian toric code and non-Abelian Ising TO\cite{Sela2014}. As mentioned earlier, this phase boundary (blue line in FIG. \ref{fig:schemaic pd}) at small field can be determined analytically as (\ref{phase bdy:ising<->toric code}) via perturbation theory (Appendix \ref{app:ising<->toric code}). As will be discussed later, symmetry analysis and topological classification dictates a quadrucritical point where all 4 phases meet. This is how we reach the phase diagram in FIG. \ref{fig:schemaic pd}.

\section{Nature of the gapless phase}

As indicated by numerical studies, in the isotropic limit ($J_\alpha\equiv J>0$), the gapless phase at an intermediate magnetic field is connected to both the non-Abelian Ising TO at small field and the polarized phase at large field, via two \emph{continuous} quantum phase transitions. This provides a strong constraint on the nature of the gapless phase. We propose this phase to be a $U(1)$ quantum spin liquid (QSL), which is described by spinon FSs coupled to a dynamical $U(1)$ gauge field at low energy. Using symmetry analysis and topological classification we show there is only one candidate $U(1)$ QSL, whose properties match the numerical results on a cylinder.

As mentioned earlier, the non-Abelian Ising TO corresponds to a chiral $p_x+\imth p_y$ TSC of spinons described by mean-field Hamiltonian (\ref{maj rep:mfh}). The universal properties of this non-Abelian phase is characterized by both the anyonic statistics of its fractionalized excitations, but also the symmetry transformation rules, known as the ``projective symmetry group'' (PSG)\cite{Wen2002,Essin2013,Barkeshli2014} of its fermionic spinons. In particular, a generic symmetry implementation on spinons has the following form in the Abrikosov-fermion representation (\ref{af rep}):
\bea\label{af rep:sym}
\Psi_i\overset{U}\longrightarrow R_U\Psi_{U(i)}G_U\big(U(i)\big).
\eea
For any symmetry group element $U\in G_s$, $R_U\in SU(2)$ is the physical spin rotations and $\{G_U(i)\in SU(2)\}$ are gauge rotations on fermionic spinons. The symmetry implementations on spinons in the Kitaev $Z_2$ QSLs are summarized in the first row of TABLE \ref{tab:psg}.

\begin{table}[tb]
\centering
\begin{tabular} {|c||c|c||c|}
\hline
State&$G_{\tilde M_h}(x_1,x_2,s)$&$G_{C_6}(x_1,x_2,s)$&Stable FS?\\
\hline
Kitaev $Z_2$&$(-1)^se^{-\imth\frac\pi4\tau_z}$&$(-1)^se^{\imth\frac{\pi}{3}\frac{\tau_x+\tau_y+\tau_z}{\sqrt3}}$&N/A\\
\hline
\hline
$U1A_{k=0}$&$e^{\imth\frac{3\pi}4\tau_z}$&$e^{-\imth\frac\pi6\tau_z}$&Yes\\
\hline
$U1B_{k=2}$&1&$\imth\tau_x\cdot e^{\imth\frac{\pi}{3}(1-2s)\tau_z}$&No\\
\hline
$U1B_{k=4}$&1&$\imth\tau_x\cdot e^{\imth\frac{\pi}{3}(2s-1)\tau_z}$&No\\
\hline
\end{tabular}
\caption{Symmetry implementations on fermionic spinons in Kitaev $Z_2$ QSLs, and in the three $U(1)$ QSLs in proximity to Kitaev $Z_2$ states. The gauge rotations for translation symmetries $T_{1,2}$ are $G_{T_{1,2}}(i)\equiv1$. The spin rotations in (\ref{af rep:sym}) associated with these symmetries are $R_{T_{1,2}}=1$, $R_{\tilde M_h}=e^{\imth\frac\pi4\sigma_z}$ and $R_{C_6}=e^{-\imth\frac{\pi}{3}\frac{\sigma_x+\sigma_y+\sigma_z}{\sqrt3}}$. (For details see Appendix \ref{app:psg})}
\label{tab:psg}
\end{table}

All symmetric $U(1)$ QSLs on honeycomb lattice preserving $T_{1,2},C_6,\tilde M_h$ symmetries (in the isotropic limit $J_\alpha=J$) can be classified by their spinon PSGs, leading to 10 distinct $U(1)$ QSLs. Numerical studies suggest a continuous phase transition between the non-Abelian Ising TO and the gapless phase, hence posing a strong constraint on the gapless phase. Among the 10 symmetric $U(1)$ QSLs, only 3 states summarized in TABLE \ref{tab:psg} are connected to the Kitaev $Z_2$ state (Ising TO) by a continuous phase transition, where spinon pairings break the emergent $U(1)$ gauge field down to $Z_2$ via the Higgs mechanism. As revealed by K-theory classification of stable Fermi surfaces (FSs), none of the three $U(1)$ QSLs in proximity to Kitaev $Z_2$ state supports robust (symmetry-protected) Dirac points in the presence of anisotropy, thus excluding the possibility of $U(1)$ Dirac spin liquids. Meanwhile only 1 state  among the 3 \ie $U1A_{k=0}$ in TABLE \ref{tab:psg} hosts stable spinon FSs (see Appendix \ref{app:psg} for details). Since a $U(1)$ QSL in 2+1-D must be stablized by gapless spinons, $U1A_{k=0}$ state becomes the only candidate for the gapless phase at intermediate field in the phase diagram.

\begin{figure}
  \includegraphics[width=\linewidth]{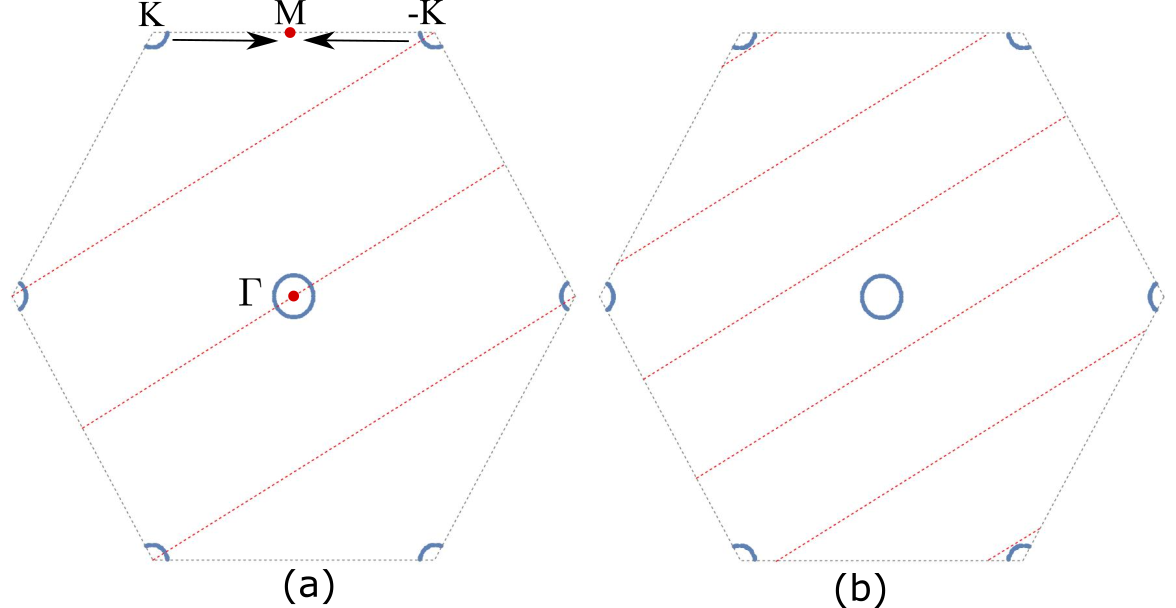}
  \caption{(Color online) How the spinon fermi surfaces (FSs) intersect with quantized momenta along the circumference of a 3-leg ladder in (a), and a 4-leg ladder in (b) (see Appendix \ref{app:U1A state mfh} for details). Blue circles denote the spinon FSs of $U1A_{k=0}$ state, including one electron pocket at $\Gamma$ and two hole pockets at $\pm K$ in the isotropic model. Red lines denote the quantized momentum along the circumference of the cylinder. Increasing the anisotropy $J_z/J_{x,y}$ not only shrinks all pockets, but also moves the two hole pockets at $\pm K$ towards $M$ point.}\label{fig:spinon fs}
\end{figure}

Furthermore, the fact that $U1A_{k=0}$ state is separated with the non-Abelian Ising phase by a continuous phase transition provides a strong constraint on its spinon FSs. In the presence of inversion symmetry (\ref{sym:inversion}), the integer-valued topological index $\nu\in\mbz$ of a gapped 2d superconductor in symmetry class D is dictated by the number of spinon FSs enclosing 4 time reversal invariant momenta (TRIM):
\bea
\nu=(\#~\text{of FSs enclosing the TRIM})\mod2
\eea
which is proved in Appendix \ref{app:topo inv}. Now that non-Abelian Ising phase corresponds to a $p_x+\imth p_y$ TSC of spinons with $\nu=1$, there must be an odd number of spinon FSs enclosing all 4 TRIM in the gapless $U1A_{k=0}$ state. As shown in FIG. \ref{fig:spinon fs}, the typical spinon FSs of an isotropic $U1A_{k=0}$ state at $J_\alpha=J$ consist of an electron pocket at zone center $\Gamma$, and one hole pocket at each zone corner $\pm K$.

%
%\begin{figure}
%		[h]
%		\parbox{4.2cm}{
%		\boxed{
%		\includegraphics[width=4cm]{h03_C1}}\\
%	(a)
%	}
%\parbox{4.2cm}{
%		\boxed{
%		\includegraphics[width=4cm]{h03_C3}}\\
%	(b)}
%%	}\\ \qquad \\ \qquad \\
%%\parbox{4.2cm}{
%%		\boxed{\includegraphics[width=4cm]{h03_C2}}\\ (c)
%%	}
%%\parbox{4.2cm}{
%%		\boxed{\includegraphics[width=4cm]{h03_C3}}\\ (d)
%%	}
%\caption{(Color online) Evolution of spinon fermi surfaces (FS) with increased [111] magnetic field $h$ and anisotropy $J_z/J$. The blue circle denotes one electron pocket around zone center $\Gamma$, and red circles denote two hole pockets around zone corners $K$ and $K^\prime$, where the 1st Brillouin zone (BZ) is a hexagon colored in black. {(a)} $h=0.25,~J_z/J=1$; {(b)} $h=0.3,~J_z/J=1$; {(c)} $h=0.3,~J_z/J=2$; {(d)} $h=0.3,~J_z/J=3$.}\label{fig:spinon fs}
%	\end{figure}

%===============Fig4:Connection to Herbertsmithite===============
\begin{figure}
  \includegraphics[width=\linewidth]{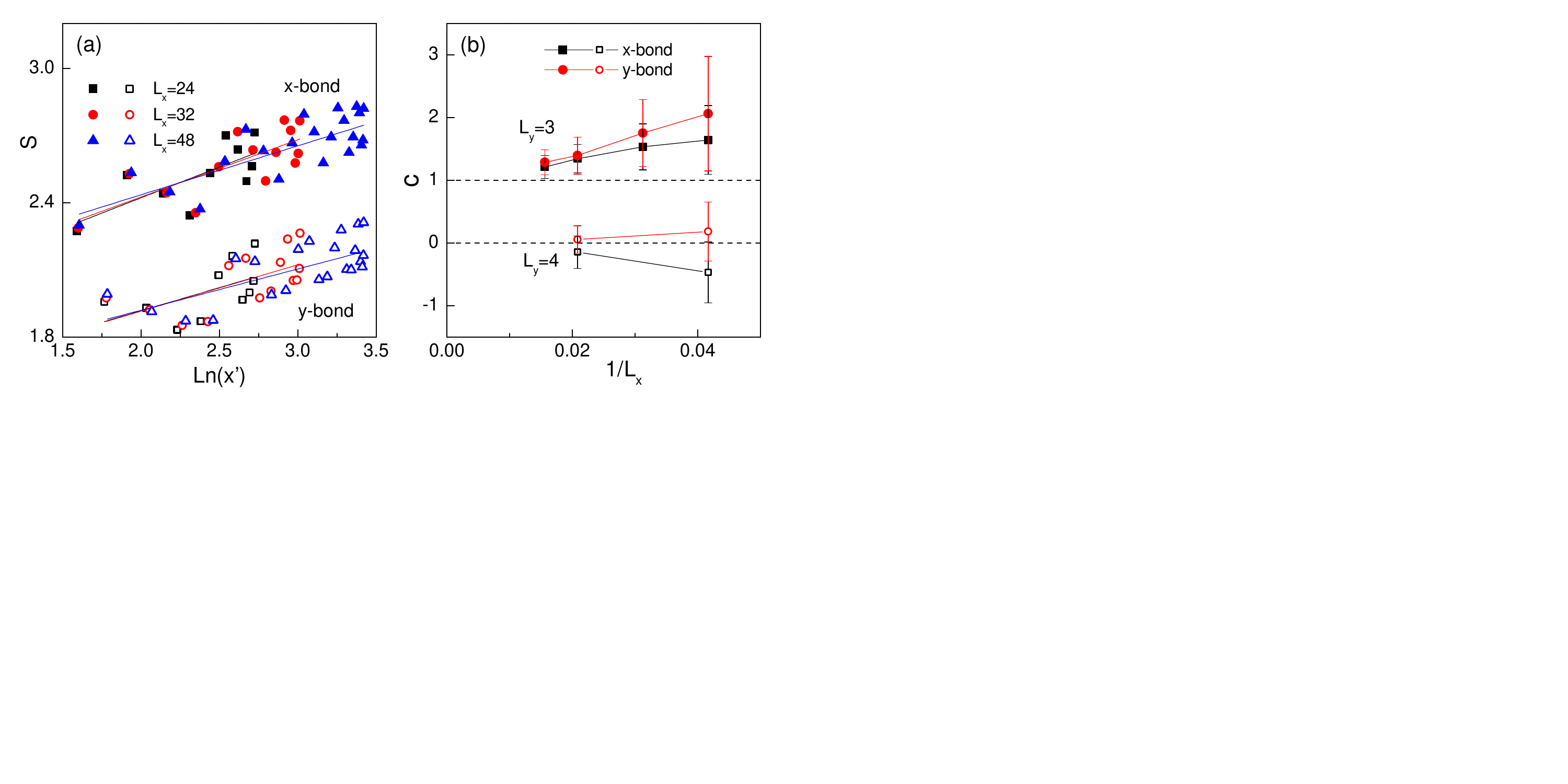}
  \caption{(Color online) (a) Von Neumann entanglement entropy $S$ on $L_y=3$ cylinders of length $L_x$, with the cut on $x$-bond and $y$-bond respectively, where $x^\prime\equiv \frac{L_x}{\pi}\sin(\frac{x\pi}{L_x}).$ (b) Extracted central charge $c$ with error bar for $L_y=3$ and $L_y=4$ cylinders. The dashed line is a guide for eyes. We choose $J_z/J=1.0$ and $h/J=0.28$ in (\ref{model}).}\label{Fig:Entropy}
\end{figure}

To further confirm the nature of the gapless $U(1)$ QSL, we use DMRG method to numerically calculate the von Neumann entanglement entropy $S=-\rm Tr(\rho ln \rho)$ on $L_x{\bf e}_x\times L_y{\bf e}_y$ cylinders of length $L_x$ and circumference $L_y$, where $\rho$ is the reduced density matrix of a subsystem with length $x$. For a 1+1-D critical system described by a conformal field theory (CFT), it is known that $S(x) =\frac{c}{6}\ln\big[\frac{L_x}{\pi}\sin(\frac{x\pi}{L_x})\big]+\tilde{c}$ on a cylinder of length $L_x$, where $c$ is the central charge of the CFT and $\tilde{c}$ is a model-dependent constant. Using this formula we extracted the central charge $c$ numerically for cylinders with circumference $L_y=3,4$, as shown in Fig.\ref{Fig:Entropy}. Here we keep up to $m=3072$ block states with a truncation error $\epsilon\leq 5\times 10^{-8}$. The 3-leg cylinder has $c\approx1$ suggesting a critical ground state, while $c\approx 0$ indicates a gapped ground state on the 4-leg cylinder. This is consistent with the spinon FSs shown in FIG. \ref{fig:spinon fs}, since the quantized momenta along the cylinder circumference only cross the two pockets at $\pm K$ for $L_y=3$ cylinder, but not $L_y=4$.

Identifying the gapless phase as $U1A_{k=0}$ state allows us to further understand the structure of the phase diagram (FIG. \ref{fig:schemaic pd}) with anisotropy. Numerical results point to a single phase of gapless $U1A_{k=0}$ state, with an odd number of spinon FSs enclosing all 4 TRIM. This suggests the neighboring gapped $Z_2$ topological order of the $U(1)$ QSL can only be $\nu=$~odd TSC of spinons, \ie the non-Abelian Ising phase. Similarly, the only gapped $Z_2$ topological order neighboring the polarized trivial phase can only be a $\nu=0$ trivial superconductor of spinons, \ie the Abelian toric code phase. This dictates the quadrucritical point joining all 4 phases, and hence the phase diagram FIG.\ref{fig:schemaic pd}.

%As one increases either anisotropy $J_z/J$ or [111] field $h/J$, the pockets shrink and ultimately vanish at the phase boundary upon entering the polarized phase. Besides, increasing anisotropy $J_z/J$ will also move the two hole pockets at $\pm K$ towards $M$, as illustrated in FIG. \ref{fig:spinon fs}.

%\bea
%\label{maj sym:T1}&(b^x,b^y,b^z,c)_{(x_1,x_2,s)}\overset{T_1}\longrightarrow(b^x,b^y,b^z,c)_{(x_1+1,x_2,s)},\\
%\label{maj sym:T2}&(b^x,b^y,b^z,c)_{(x_1,x_2,s)}\overset{T_2}\longrightarrow(b^x,b^y,b^z,c)_{(x_1,x_2+1,s)},\\
%\label{maj sym:mirror}&(b^x,b^y,b^z,c)_{(x_1,x_2,s)}\overset{\tilde M_h}\longrightarrow(-1)^s(b^y,b^x,b^z,-c)_{(x_2,x_1,s)},\\
%\label{maj sym:c6}&(b^x,b^y,b^z,c)_{(x_1,x_2,s)}\overset{C_6}\longrightarrow (-1)^s(b^z,b^x,b^y,c)_{(1-x_2,x_1+x_2-s,1-s)}.
%\eea

\section{Quantum phase transitions}

In this section we discuss quantum phase transitions between the 4 distinct quantum phases in FIG. \ref{fig:schemaic pd}. The strategy is to start from the quadrucritical point at the intersection of 4 phase boundaries, then to extend to the 4 phase boundaries. The low-energy physics of the quadrucritical point is described by the following effective field theory
\bea\label{qcp theory}
\mathcal{L}_\text{quadru}=~~~~~~~~~~~~~~~~~~~~~~~~~~~~~~~~~~~~~~~~~~~~~~~~~~~\\
\notag\psi^\dagger_0(\imth\partial_t+a_0+\mu\frac{m_1}{m_0}+\mu_0|\Delta|^2)\psi_0-\psi^\dagger_0\frac{(-\imth\vec\nabla-\vec a)^2}{2m_0}\psi_0\\
\notag+\psi^\dagger_1(\imth\partial_t+a_0-\mu+\mu_1|\Delta|^2)\psi_1+\psi^\dagger_1\frac{(-\imth\vec\nabla-\vec a)^2}{2m_1}\psi_1\\
\notag+\Delta({\bf r})\sum_{\alpha=0,1}\psi_\alpha(-\imth v_{x\alpha}\partial_x- v_{y\alpha}\partial_y)\psi_\alpha+h.c.\\
-\frac1{4g}f_{\mu\nu}f^{\mu\nu}-\frac{\rho}2|(-\imth\vec\nabla+2\vec a)\Delta|^2-\frac t2|\Delta|^2-u|\Delta|^4+\cdots\notag
\eea
Here $\psi_0$ denotes an electron-like spinon FS at zone center $\Gamma$, while $\psi_1$ denotes a hole-like spinon FS at hexagonal BZ edge center $M$ with $k_1=k_2=\pi$. In the absence of spinon pairing $|\Delta|=0$ \ie when $t>0$ in (\ref{qcp theory}), since fermionic spinons $\{f_{i,\sigma}\}$ have an integer filling of 2 spinons per unit cell, due to Luttinger's theorem the ``electron density'' at $\Gamma$ must equal the ``hole density'' at $M$:
\bea
\frac1{\mathcal{V}}\int\text{d}^2{\bf r}\psi_1\psi^\dagger_1=\frac1{\mathcal{V}}\int\text{d}^2{\bf r}\psi^\dagger_0\psi_0=\frac{m_1\mu}{2\pi}
\eea
However with spinon pairing $|\Delta|>0$ \ie when $t<0$ in (\ref{qcp theory}), Luttinger's theorem is violated, as captured by the $\mu_{0,1}>0$ terms in (\ref{qcp theory}). While the bottom of the ``electron'' pocket at $\Gamma$ crosses zero energy at $\mu=-\frac{\mu_0m_0}{m_1}|\Delta|^2=\frac{\mu_0m_0}{m_1u}t\cdot\theta(-t)\leq0$, the top of the ``hole'' pocket at $M$ crosses zero energy when
\bea
\mu=\mu_g|\Delta|^2=-\frac{\mu_1}{u}t\cdot\theta(-t)\geq0
\eea
where $\theta(x)$ stands for the step function. As shown in Appendix \ref{app:topo inv}, guaranteed by inversion symmetry (\ref{sym:inversion}), the number of spinon FSs surrounding 4 time reversal invariant momenta (TRIM) including $\Gamma$ and $M$ implies the topology of the consequent gapped superconducting state: an odd number of spinon FSs necessarily leads to a chiral TSC. Therefore the 4 distinct phases are captured in effective theory (\ref{qcp theory}) as

(1) Gapped non-Abelian Ising phase:
\bea
t<0,~~\mu>\mu_1|\Delta|^2=-\frac{\mu_1}{u}t
\eea
(2) Abelian toric code phase:
\bea
t<0,~~\mu<\mu_1|\Delta|^2=-\frac{\mu_1}{u}t
\eea
(3) Gapless $U(1)$ QSL with spinon fermi surfaces:
\bea
t>0,~~\mu>0
\eea
(4) Gapped spin polarized (trivial) phase:
\bea
t>0,~~\mu<0
\eea
Now we can comment on the 4 phase boundaries in FIG.\ref{fig:schemaic pd}. The blue phase boundary $\mu=-\frac{\mu_1}{u}t,~t<0$ between the two gapped topological orders is driven by the ``band inversion'' at $M$ in the presence of spinon pairings, consistent with the exact solution at zero field with $J_z=2J_{x,y}$ (see Appendix \ref{app:ising<->toric code}). The critical theory across this blue phase boundary is described by a single Majorana fermion coupled to a dynamical gauge field. The upper phase boundary $t=0,~\mu<0$ of the red line separates the polarized phase and Abelian toric code, driven by ``strong pairing'' of spinons in the absence of spinon FS. The lower phase boundary of the red line $\mu=0,~t>0$ is a metal-insulator transition of spinons coupled to $U(1)$ gauge fields, driven by vanishing the spinon FS at both $\Gamma$ and $M$ in the absence of pairing. Finally, the black phase boundary $t=0,~\mu>0$ describes a transition between a gapless $U(1)$ QSL with spinon FSs and gapped non-Abelian Ising phase, driven by the ``weak pairing''\cite{Read2000} of spinons on the fermi surface. Notice that along both the black and lower part of red phase boundaries, as anisotropy $J_z/J$ is increased, the two hole-like pockets at $\pm K$ will move towards $M$ and ultimately merge into one hole pocket at $M$.

\section{Summary}

To conclude, through a combination of symmetry analysis, topological classification and numerical studies, we obtain a phase diagram of Kitaev model as a function of bond anisotropy $J_z/J_{x,y}$ and perpendicular magnetic field $h_{[111]}$. We identify 4 distinct phases separated from each other by continuous quantum phase transitions, where four continuous phase boundaries intersect at a quadrucritical point as dictated by inversion symmetry. In particular, we identify a symmetric gapless $U(1)$ QSL with spinon fermi surfaces, $U1A_{k=0}$ state, as the only candidate of the intermediate phase between non-Abelian Ising phase and polarized trivial phase in the isotropic Kitaev model under a [111] field. While at this stage it is difficult to directly relate our theoretical results to the experimental observations in $\alpha$-RuCl$_3$, this work paved the road for future efforts on various QSL phase and quantum critical phenomena in Kitaev materials.

\acknowledgements{YML thanks Lesik Motrunich, Ying Ran, Yong-Baek Kim and Nandini Trivedi for feedbacks, Yin-Chen He for sharing unpublished numerical results, and Aspen Center for Physics for hospitality where part of the manuscript was written. HCJ thanks Yifan Jiang for insightful discussions. This work is supported by NSF under award number DMR-1653769 (CYW,YML), by U.S. ARO (W911NF11-1-0230), AFOSR (FA9550-16-1-0006), MURI-ARO (W911NF-17-1-0323) (BH) , by the Department of Energy, Office of Science, Basic Energy Sciences, Materials Sciences and Engineering Division, under Contract DE-AC02-76SF00515 (HCJ), and in part by NSF grant PHY-1607611 (YML).}

\emph{Note:} After completion of this work, we became aware of an independent work by Yin-Chen He and others, who studied the isotropic Kitaev model under magnetic field using iDMRG.

\bibliographystyle{apsrev4-1}
\bibliography{bibs_career}

\begin{widetext}

\appendix

\section{Phase boundary near $J_z=2J$}\label{app:ising<->toric code}

At zero field, Kitaev had shown\cite{Kitaev2006} that model (\ref{model}) with $h=0$ hosts a gapless $Z_2$ spin liquid (``phase B'') ground state as long as $|J_z|<|J_x|+|J_y|=2J$. Meanwhile $J_z>2J$ leads to a gapped $Z_2$ spin liquid (``phase A'') with an Abelian topological order of the toric code type. A small magnetic field $\vec h\parallel(1,1,1)$ will drive the gapless phase B into a gapped spin liquid with a non-Abelian topological order of the Ising type. These facts indicate a phase boundary separating the Abelian toric code (gapped phase A) and the non-Abelian Ising phase at a finite (but small) perpendicular field, by tuning the anisotropy parameter $J_z/J$. Below we derive this phase boundary around $J_z=2J$ and $h=0$.

\subsection{Kitaev's exact solution}

The zero-field Kitaev model (\ref{model}) can be solved exactly by the following Majorana representation:
\bea\label{majorana rep}
2\hat S^\alpha=\hat\sigma^\alpha=\imth b^\alpha c,~~~\alpha=x,y,z.
\eea
where $\vec\sigma$ are 3 Pauli matrices, and $\{b^\alpha,c\}$ are 4 Majorana fermions representing the spin-$1/2$ d.o.f. per site. To faithfully reproduce the spin-$1/2$ Hilbert space, the following constraint must be implemented on each site:
\bea\label{constraint}
b^x_ib_i^yb_i^zc_i=1,~~~\forall~i.
\eea
and therefore
\bea\label{majorana rep full}
2\hat S^\alpha=\imth b^\alpha c=-\imth\frac{\epsilon_{\alpha\beta\gamma}}2b^\beta b^\gamma.
\eea
At zero field, model (\ref{model}) can be rewritten in terms of Majorana fermions as
\bea\label{model:zero field}
\hat H_0\equiv\hat H_K(h=0)=-\frac14\sum_{\langle i,j\rangle}(\imth J_{\alpha_{ij}}b_i^{\alpha_{ij}}b_j^{\alpha_{ij}})(\imth c_ic_j)
\eea
which describes how Majorana fermions $\{c_i\}$ hop with amplitude $t_{ij}=\imth J_{\alpha_{ij}}b_i^{\alpha_{ij}}b_j^{\alpha_{ij}}$. As pointed out by Kitaev, the $Z_2$ flux around each hexagonal plaquette $p$ is a conserved quantity of model (\ref{model}) at zero field:
\bea\label{z2 flux}
W_p=\prod_{\langle i,j\rangle\in p}(\imth b_i^{\alpha_{ij}}b_j^{\alpha_{ij}})=\prod_{i\in p,k\notin p}\sigma_i^{\alpha_{ik}}=\pm1.
\eea
Lieb's theorem\cite{Lieb1994} indicates that the lowest energy (ground) state has a uniform zero flux of $W_p=+1, \forall~p$. In the zero-flux sector, one can choose a gauge where
\bea
b_i^{\alpha_{ij}}b_j^{\alpha_{ij}}\equiv\imth,~~~i\in\text{A sublattice},~j\in\text{B sublattice}.
\eea
While $\{b_i^\alpha\}$ fermions pairwise form a dimer on each link, $\{c_i\}$ fermions can hop in the background $Z_2$ flux (\ref{z2 flux}). The BdG band structure of Majorana fermions $\{c_j\}$ in momentum space from (\ref{model:zero field}) can be written as
\bea\label{fermion spectrum:no field}
&H^{(0)}_{\bf k}(h=0)=\frac{\imth J}4\bpm&-f_{\bf k}\\f^\ast_{\bf k}&\epm,\\
&\notag f_{\bf k}=\frac{J_z}{J}+e^{\imth k_1}+e^{\imth k_2}
\eea
in the basis of $(c_{A,{\bf k}},c_{B,{\bf k}})^T$ where $A,B$ labels the two sublattices.

Clearly the $\{c_i\}$ spectrum is fully gapped (\ie $|f_{\bf k}|>0,~\forall~{\bf k}$) if and only if $|J_z/J|>2$. This corresponds to the ``gapped A phase'', with an Abelian topological order of the toric code type. On the other hand if $|J_z/J|<2$, there will be a pair Dirac points at zero energy, located at momenta $k_1=-k_2=\pm\arccos\frac{-J_z}{2J}$. This corresponds to the ``gapless B phase'', described by Dirac fermions coupled to dynamical $Z_2$ gauge fields in the low energy limit.

\subsection{Perturbation theory in the zero flux sector}

As shown by Kitaev\cite{Kitaev2006}, above the low-energy states within the zero flux sector with $W_p\equiv+1,~\forall~p$, there is a finite energy gap $\Delta$ ($\sim0.27|J|$ in the isotropic $J_\alpha\equiv J$ case) for each $\pi$ flux excitation (\ie $W_p=-1$). Within the low energy sector of zero flux states, below we summarize the effects of magnetic field $\vec h$ up to 3rd order in perturbation theory.

Denoting the projector into zero-flux Hilbert space of (\ref{model:zero field}) as $\Pi_0$, the perturbation expansion can be written as
\bea
&\notag H=H_0+H_1+H_2+\cdots,\\
&H_n=\Pi_0\big(\hat V\frac{1-\Pi_0}{E_0-\hat H_0}\big)^{n-1}\hat V\Pi_0,~~~\forall~n\geq1.\label{perturbation expansion}
\eea
where $\hat V$ is the perturbation term. In our case, $\hat H_0$ is given by (\ref{model:zero field}) while
\bea
\hat V=-\vec h\cdot\sum_{i}{\bf S}_i,~~~\vec h=h(1,1,1).
\eea
and the ground states projector is written as
\bea
\hat \Pi_0=\prod_p\big(\frac{1+\hat W_p}2\big)
\eea

It's straightforward to show that 1st order perturbation theory vanishes, while the 2nd order terms renormalize the original Kitaev terms between each NN
\bea\label{2nd order}
\hat H_2=-{2h^2}\sum_{\langle i,j\rangle}S_i^{\alpha_{ij}}S_j^{\alpha_{ij}}/\Delta_{2,\alpha_{ij}}
\eea
where $\Delta_{2,\alpha}$ is the energy cost for a pair of flux excitations sharing a NN link along $\alpha$ direction. At anisotropic coupling $J_{x,y}=J$ and $J_z=2J$, explicit calculations show that
\bea
\Delta_{2,x}=\Delta_{2,y}\approx0.063J,~~~\Delta_{2,z}\approx0.078J.
\eea

Following Kitaev, there are two types of 3rd order contributions in the perturbation expansion:
\bea
\hat H_3=-g_{3}\frac{h^3}{J^2}\Big(\sum_{\langle ik\rangle \langle jk\rangle}S_i^{\alpha_{ik}}S_j^{\alpha_{jk}}S_k^{\beta\neq\alpha_{ik},\alpha_{jk}}+\sum_{\langle il\rangle \langle jl\rangle\langle kl\rangle}S_i^{\alpha_{il}}S_j^{\alpha_{jl}}S_k^{\alpha_{kl}}\Big)\label{3rd order}
\eea
where the coupling constant $g_3$ is given by
\bea
g_3=\frac{4J^2}{\Delta_{2,z}\Delta_{2,x/y}}+2\big(\frac{J}{\Delta_{2,z}}\big)^2.
\eea
In the Majorana fermion representation (\ref{majorana rep}), the 3rd order terms can be written as
\bea
&\notag\hat H_3=-\frac{g_{3}}8\frac{h^3}{J^2}\Big[\sum_{\langle ik\rangle \langle jk\rangle}\epsilon_{\alpha_{ik}\beta\alpha_{jk}}\cdot(\imth b_i^{\alpha_{ik}}b_k^{\alpha_{ik}})(\imth b_j^{\alpha_{jk}}b_k^{\alpha_{jk}})(\imth c_ic_j)\\
&+\sum_{\langle il\rangle \langle jl\rangle\langle kl\rangle}\epsilon_{\alpha_{il}\alpha_{jl}\alpha_{kl}}\cdot(\imth b_i^{\alpha_{il}}b_l^{\alpha_{il}})(\imth b_j^{\alpha_{jl}}b_l^{\alpha_{jl}})(\imth b_k^{\alpha_{kl}}b_l^{\alpha_{kl}})c_ic_jc_kc_l\Big]
\eea
Clearly the 1st term in (\ref{3rd order}) introduces a 2nd NN hopping between Majoranas $\{c_i\}$, while the 2nd term in (\ref{3rd order}) becomes 4-fermion interactions between Majorana fermions $\{c_i\}$.

%We do not exhaust all 4th order terms in the perturbation expansion due to its complexity, but only focus on those that renormalize the Kitaev NN interactions. In particular, we have
%

\subsection{Determining the phase boundary}

Therefore up to 3rd order terms in the perturbation expansion, within the low-energy Hilbert space of zero flux sector, the effective Hamiltonian for $\{c_i\}$ fermions are the following
\bea\label{mfh:3rd order}
&\hat H_{eff}=\frac{J}{4}\bpm c_{-{\bf k},A}\\c_{-{\bf k},B}\epm^T\bpm G_{\bf k}&-\imth F_{\bf k}\\ \imth F^\ast_{\bf k}&-G_{\bf k}\epm\bpm c_{{\bf k},A}\\c_{{\bf k},B}\epm\\
\notag&-\frac{g_3h^3}{8J^2}\sum_{\langle il\rangle \langle jl\rangle\langle kl\rangle}(-1)^l\epsilon_{\alpha_{il}\alpha_{jl}\alpha_{kl}}c_ic_jc_kc_l+O(\frac{h^4}{J^3}).
\eea
where the matrix elements are given by
\bea
\notag&F_{\bf k}=\frac{J_z}{J}-\frac{2h^2}{J\Delta_{2,z}}+(1-\frac{2h^2}{J\Delta_{2,x/y}})(e^{\imth k_1}+e^{\imth k_2}),\\
&G_{\bf k}=\frac{g_3}{4}(\frac hJ)^3\big[-\sin k_1+\sin k_2+\sin(k_1-k_2)\big].~~~
\eea
It's straightforward to show that as long as $|F_{\bf k}|>0,~\forall~{\bf k}$, the system will have a gapped ground state with Abelian topological order of toric code type. On the other hand, if $F_{\bf k}$ vanishes at certain momenta leading to Dirac fermions, $G_{\bf k}$ will gap out the Dirac fermions and give rise to a non-Abelian topological order of Ising type. The short-range 4-fermion interaction is irrelevant for the Dirac fermions and hence does not modify the phase.

As a result, the phase boundary between Abelian toric code and non-Abelian Ising phases is given by the following condition:
\bea
\frac{J_z}{J}-\frac{2h^2}{J\Delta_{2,z}}=2(1-\frac{2h^2}{J\Delta_{2,x/y}})+O(\frac{h}{J})^4
\eea
which determines whether $|F_{\bf k}|>0,~\forall~{\bf k}$ or not. This leads to the following phase boundary around $J_z=2J$ at small fields:
\bea
\frac{J_z}{J}\approx2-38(\frac{h}{J})^2+O(\frac hJ)^4.
\eea

\section{Phase boundary in the anisotropic limit}\label{app:polarized<->toric code}

In the strongly anisotropy limit $J_z\gg J$, a perturbation expansion in terms of $J/J_z\ll1$ can be performed. In the absence of field, Kitaev showed that the leading term in this strong anisotropy expansion exactly corresponds to the toric code Hamiltonian. Here we show that with the external field along $(1,1,1)$ direction, the leading terms in the perturbation expansion correspond to the toric code of coupling strength $\sim J^4/J_z^3$ under a transverse field of strength $\sim h^2/J_z$. As was shown in Ref.\cite{Vidal2009,Vidal2009a,Dusuel2011}, a transverse field comparable to the toric code terms will drive the Abelian $Z_2$ topological order into a confined trivial phase. Therefore this allows us to determine the phase boundary between toric code phase and spin-polarized trivial field, in the limit of small field and strong anisotropy \ie $J/J_z,h/J_z\ll1$.

\subsection{Perturbation theory in the strong anisotropy limit}

Following Kitaev\cite{Kitaev2006}, in the strong anisotropy limit of $J_z\gg J=J_{x,y}$, we perform perturbation expansion in terms of $J/J_z$ and $h/J_z$. The 0-th order Hamiltonian is
\bea
\hat H_0=J_z\sum_{\alpha_{ij}=z}S_i^zS_j^z-h_z\sum_i S_i^z.
\eea
and the perturbation is given by
\bea
\notag\hat V=\hat H_K(h)-\hat H_0=\\
J\sum_{\alpha_{ij}=x,y}S_i^{\alpha_{ij}}S_j^{\alpha_{ij}}-\sum_i(h_x S_i^x+h_y S_i^y)
\eea
We discuss the perturbation theory for a small field $\vec h=(h_x,h_y,h_z)$ along any direction, but in the end restrict ourselves to the case of model (\ref{model}) with
\bea
h_x=h_y=h_z=h.
\eea

The ground state manifold is given by the following constraint on each pair of NN spins on a $z$-link $l\equiv\langle i,j\rangle$:
\bea
S_i^z=-S_j^z=\pm\frac12,~~~\forall~\alpha_{ij}=z.
\eea
Therefore we label the two states of the ``block spin'' in each unit cell as
\bea
&\dket{Z_l=\pm1}\equiv\dket{S_i^z=\pm\frac12,S_j^z=\mp\frac12},\\
&\notag l\equiv\langle ij\rangle~\forall~\alpha_{ij}=z,~~~i\in A,~j\in B.
\eea
The 3 Pauli matrices for the block spin $l$ on NN link $\langle ij\rangle$ with $\alpha_{ij}=z$ are given by
\bea
&Z_l=\sigma_i^z=-\sigma_j^z,~~~i\in A,~j\in B,\\
&X_l=\sigma_i^x\sigma_j^x=\sigma_i^y\sigma_j^y,\\
&Y_l=\sigma_i^y\sigma_j^x=-\sigma_i^x\sigma_j^y.
\eea
since $\sigma_i^z\sigma_j^z=-1$ for $\vec\sigma_i=2{\bf S}_i$.

To create excitations beyond state manifold, there is an excitation gap of
\bea
\Delta_\pm\equiv\frac{J_z}2\pm h_z=\frac{J_z}2\pm h
\eea
to flip one single spin on a NN z-bond, where $\pm$ sign corresponds to A/B sublattices.

In the perturbation expansion (\ref{perturbation expansion}), the first order term vanishes while the 2nd order term is given by
\bea
&\hat H_2=-B_x\sum_lX_l,\\
\notag&B_x={\big[(\frac{h_x}2)^2+(\frac{h_y}2)^2\big]}(\frac1{\Delta_+}+\frac1{\Delta_-})=\frac{2h^2}{J_z}+O(\frac{h^3}{(J_z)^2})
\eea
The 3rd order terms are
\bea
\notag\hat H_3=\frac{J}{J_z^2}\Big\{\sum_{q=p+\vec a_1}\big[\big(h_x(X_p+1)+h_y Y_p\big)\cdot\big(h_x(X_q+1)+h_yY_q\big)-h_y^2Z_pZ_q\big]+\\
\sum_{q=p+\vec a_2}\big[\big(h_y(X_p+1)-h_x Y_p\big)\big(h_y(X_q+1)+h_xY_q\big)-h_x^2Z_pZ_q\big]\Big\}+O\big(J(\frac h{J_z})^3\big).~~~~
\eea
including both bilinear spin interactions between NNs and transverse fields. The coupling constant of 3rd order terms is in the order of $g_3\sim h^2J/J_z^2$.

Finally the 4th order terms in the perturbation expansion lead to the toric code Hamiltonian
\bea
\hat H_4=-g_4\sum_{\langle uldr\rangle}Y_lY_rZ_uZ_d+O(\frac{J^2h^2}{J_z^3}),\\
\notag g_4=\frac{J^4}{64J_z^3}.
\eea
where $\langle uldr\rangle$ labels 4 NN block spins in a diamond pattern.

\subsection{Determining the phase boundary}

Below we consider the perturbation theory in the following small field, anisotropy limit:
\bea\label{strong anisotropy+small field}
\frac hJ\ll\sqrt{\frac J{J_z}}\ll1.
\eea
Clearly in this limit, we have
\bea
\frac{g_3}{g_4}\sim\frac{h^2J_z}{J^3}\ll1
\eea
Hence the effective Hamiltonian in the ground state manifold from perturbation theory is dominated by the toric code model under a transverse field:
\bea\label{toric code+field}
\hat H_{eff}=-g_4\sum_{\langle uldr\rangle}Y_lY_rZ_uZ_d-B_x\sum_lX_l+\cdots,\\
\notag B_x=\frac{2h^2}{J_z},~~g_4=\frac{J^4}{64J_z^3}.~~~~~
\eea
where $\cdots$ stands for all other terms, with much smaller couplings compared to the 2 terms above.

As shown in Ref.\onlinecite{Vidal2009,Vidal2009a,Dusuel2011}, model (\ref{toric code+field}) exhibits 2 different phases: the Abelian toric code phase with anyons and the confined spin-polarized phase without anyons, and their phase boundary is given by $g_4\simeq B_x$. Therefore in the small field and strong anisotropy limit (\ref{strong anisotropy+small field}), the phase boundary between Abelian $Z_2$ topological order and spin-polarized phase is given by
\bea
g_4\simeq B_x\Longleftrightarrow\frac{J_z}J=C_0(\frac hJ)^{-1}\gg1
\eea
where $C_0$ is a constant of order 1.

It's also straightforward to show that upon increasing the magnetic field $h_{[111]}$, the confined phase of transverse-field toric code model (\ref{toric code+field}) with
\bea
-\sigma_i^z\sigma_j^z=\sigma_i^x\sigma_j^x=1,~~~\forall~\alpha_{ij}=z.
\eea
is adiabatically connected to (without phase transitions) the polarized phase, where all spins point to [111] direction. This suggests a single quantum phase transition when increasing magnetic field in the strong anisotropy limit $J_z/J\gg1$.

\section{$U(1)$ spin liquids in proximity to the Kitaev state}\label{app:psg}

In both the Abelian toric-code phase and the non-Abelian Ising phase, there is one type of anyons (or one superselection sector) obeying fermi statistics, coined ``spinons'' here. They are nothing but the Majorana fermions in Kitaev's exact solution, which form a strong-pairing trivial superconductor in the toric code phase, or a weak-pairing $p+\imth p$ superconductor with chiral Majorana edge modes in the Ising phase. If these fermionic spinons instead form a fermi liquid, it corresponds to a $U(1)$ spin liquid with an emergent spinon fermi surface. Below we first examine the symmetry implementation on these fermionic spinons in both toric code and Ising phases. In a $U(1)$ spin liquid connected to the Ising phase by a continuous quantum phase transition, the symmetry implementations on spinons must be compatible with those in the Ising phase. This principle allows us to classify all possible $U(1)$ spin liquids in proximity to Kiteav's $Z_2$ states, and identify a promising candidate for the field induced gapless spin liquid.

\subsection{Fermion symmetry fractionalization in the Kitaev $Z_2$ spin liquidss}

As shown previously in (\ref{model:zero field}) and (\ref{mfh:3rd order}), the ground state of Kitaev model under a small field can be solved exactly using the Majorana fermion representation (\ref{majorana rep full}) of spin-$\frac12$'s. The ground state $\dket{g.s.}$ is given by Gutzwiller projection on the spinon mean-field state $\dket{MF}$:
\bea
\dket{g.s.}=\prod_{i}(1-b_i^xb_i^yb_i^zc_i)\dket{MF}
\eea
where $\dket{MF}$ is the ground state of the following mean-field Hamiltonian for Majorana fermions
\bea
\notag\hat H_{MF}^{Z_2}=-\sum_{\langle i\in A,j\in B\rangle}\big(\frac{\Delta_{2,\alpha_{ij}}}2\imth b_i^{\alpha_{ij}}b_j^{\alpha_{ij}}+\frac{\tilde J_{\alpha_{ij}}}4\imth c_ic_j\big)\\
\notag-\imth h\sum_i(b_i^x+b_i^y+b_i^z)c_i~~~~~\\
\label{mfh:kitaev model}+\frac{g_3h^3}{8J^2}\sum_{\langle\langle i,j\rangle\rangle}\nu_{ij}\imth c_ic_j+O(\frac{h^3}{J^2})~~~~~
\eea
where we defined the renormalized NN couplings as
\bea
\tilde J_{\alpha_{ij}}\equiv J_{\alpha_{ij}}-\frac{2h^2}{J\Delta_{2,\alpha_{ij}}}.
\eea

In the isotropic case ($J_{\alpha}\equiv J,~\Delta_{2,\alpha}\equiv\Delta_2$), the spinon mean-field state preserves translations $T_{1,2}$, rotation $C_6$ and anti-unitary ``magnetic mirror'' $\tilde M_h=M_{[1\bar10]\cdot\bst}$. The whole symmetry group $G_s$ of isotropic model (\ref{model}) is given by
\bea\label{sym group}
G_s=\{T_1^{\nu_1}T_2^{\nu_2}\tilde M_h^{\nu_m}C_6^{\nu_6}|\nu_{1,2}\in\mbz,\nu_m\in\mbz_2,\nu_6\in\mbz_6\}.~~~~
\eea
Under these symmetries, the Majorana fermions transform as
\bea
\label{sym:maj:T1}&\bpm b^x\\b^y\\b^z\\c\epm_{(x_1,x_2,s)}\overset{T_1}\longrightarrow\bpm b^x\\b^y\\b^z\\c\epm_{(x_1+1,x_2,s)},\\
\label{sym:maj:T2}&\bpm b^x\\b^y\\b^z\\c\epm_{(x_1,x_2,s)}\overset{T_2}\longrightarrow\bpm b^x\\b^y\\b^z\\c\epm_{(x_1,x_2+1,s)},\\
\label{sym:maj:mirror}&\bpm b^x\\b^y\\b^z\\c\epm_{(x_1,x_2,s)}\overset{\tilde M_h}\longrightarrow(-1)^s\bpm b^y\\b^x\\b^z\\-c\epm_{(x_2,x_1,s)},\\
\label{sym:maj:c6}&\bpm b^x\\b^y\\b^z\\c\epm_{(x_1,x_2,s)}\overset{C_6}\longrightarrow (-1)^s\bpm b^z\\b^x\\b^y\\c\epm_{(1-x_2,x_1+x_2-s,1-s)}.
\eea

The above symmetry implementations have the following algebra when acting on fermionic spinons $\{b^\alpha_i\}$
\bea\label{psg:T1,T2}
&T_1T_2T_1^{-1}T_2^{-1}=1,\\
\label{psg:m,T1}&\tilde M_h^{-1}T_1\tilde M_hT_2^{-1}=1,\\
\label{psg:m,T2}&\tilde M_h^{-1}T_2\tilde M_hT_1^{-1}=1,\\
\label{psg:c6,T1}&C_6^{-1}T_1C_6T_2T_1^{-1}=1,\\
\label{psg:c6,T2}&C_6^{-1}T_2C_6T_1^{-1}=1,\\
\label{psg:m}&(\tilde M_h)^2=1,\\
\label{psg:c6}&(C_6)^6=(-1)^{\hat F},\\
&(C_6\tilde M_h)^2=1.\label{psg:c6,m}
\eea
where
\bea
(-1)^{\hat F}=\prod_i(b_i^xb_i^yb_i^zc_i).
\eea
is the total number parity of fermionic spinons. The above algebra hold for the Abelian toric code phase (\ie the trivial strong-pairing superconductor of fermionic spinons) and the non-Abelian Ising phase (\ie the weak-pairing $p+\imth p$ chiral topological superconductor of fermionic spinons), both of which host one type of bulk anyon (or superselection sector) $\epsilon$ obeying fermi statistics. The algebraic relations (\ref{psg:T1,T2})-(\ref{psg:c6,m}) characterize the projective symmetry group (PSG)\cite{Wen2002} of fermionic spinons $\epsilon$, as the mathematical description for symmetry fractionalization of fermionic spinon $\epsilon$ in both toric code and Ising topological orders.

Finally, we discuss the relation between Majorana representation (\ref{majorana rep full}) and the more familiar Abrikosov representation of spin-$1/2$'s. In the Abrokosov fermion representation, each spin-$1/2$ is represented by a pair of complex fermions $\{f_{i\uparrow},f_{i\downarrow}\}$ as follows
\bea
{\bf S}_i=\frac14\text{Tr}\big(\Psi^\dagger_i\vec{\sigma}\Psi_i\big)
\eea
where we have defined spinon operator
\bea
\Psi_i=\bpm f_{i\uparrow}&f^\dagger_{i\downarrow}\\f_{i\downarrow}&-f_{i\uparrow}^\dagger\epm=(\imth\sigma_y)\Psi^\ast_i(\imth\tau_y)
\eea
It has a one-to-one correspondence with the Majorana fermion representation (\ref{majorana rep full}) given by the following relation
\bea
f_{i\uparrow}=\frac{b_i^z+\imth c_i}2,~~f_{i\downarrow}=\frac{b_i^x+\imth b_i^y}2,~~~\\
\Psi_i=\bpm f_{i\uparrow}&f^\dagger_{i\downarrow}\\f_{i\downarrow}&-f_{i\uparrow}^\dagger\epm=\frac12\big(\sum_{\alpha=x,y,z}b_i^\alpha\hat\sigma_\alpha+\imth c~\hat 1_{2\times2}\big).~~~
\eea
where the single-occupancy constraint per site for complex fermions $\{f_{i\uparrow},f_{i\downarrow}\}$ is nothing but constraint (\ref{constraint}) for Majorana fermions.

Under a symmetry operation $U$, the complex fermions transform as
\bea\label{sym:abk general}
\Psi_i\overset{U}\longrightarrow R_U\Psi_{U(i)}G_U\big(U(i)\big).
\eea
where $R_U\in SU(2)$ corresponds to physical spin rotations and $\{G_U(i)\in SU(2)\}$ corresponds to the gauge rotations associated with symmetry operation $U$. It is straightforward for any unitary symmetry, but has some subtlety for an anti-unitary symmetry. Take the familiar time reversal symmetry $\bst$ of spin-$1/2$ fermions for example, we have
\bea
\Psi_i\overset{\bst}\longrightarrow\imth\sigma_y\Psi_i=-\Psi_i^\ast(\imth\tau_y).
\eea
Therefore under magnetic mirror symmetry $\tilde M_h\equiv M_{[1\bar10]}\cdot\bst$ it transforms as
\bea
\Psi_i\overset{\tilde M_h}\longrightarrow\imth\sigma_yU_{M_{[1\bar10]}}\Psi_{\tilde M_h(i)} G_{\tilde M_h}\big(\tilde M_h(i)\big),\\
\notag U_{M_{[1\bar10]}}=\imth\frac{\sigma_x-\sigma_y}{\sqrt2}.~~~~~
\eea

In the Kitaev model (\ref{model}) under a $\langle111\rangle$ magnetic field, the physical spin rotations are
\bea
R_{T_{1}}=R_{T_2}=1,\\
R_{\tilde M_h}=\imth\sigma_y\cdot\frac{\imth(\sigma_x-\sigma_y)}{\sqrt2}=e^{\imth\frac\pi4\sigma_z},\\
R_{C_6}=e^{-\imth\frac{\pi}{3}\frac{\sigma_x+\sigma_y+\sigma_z}{\sqrt3}}.
\eea
where $\bst=\imth\sigma_y\cdot\mathcal{K}$ is the time reversal operation and $\mathcal{K}$ represents complex conjugation.

In the Kitaev $Z_2$ spin liquids described by (\ref{mfh:kitaev model}), the gauge rotations associated with the symmetry operations are given by
\bea\label{sym:abk:T1,2}
G_{T_1}(i)=G_{T_2}(i)=1,~~~~~\forall~i;\\
\label{sym:abk:mirror}G_{\tilde M_h}(x_1,x_2,s)=(-1)^s\cdot e^{-\imth\frac\pi4\tau_z};\\
\label{sym:abk:c6}G_{C_6}(x_1,x_2,s)=(-1)^s\cdot e^{\imth\frac{\pi}{3}\frac{\tau_x+\tau_y+\tau_z}{\sqrt3}}.
\eea
where we use Pauli matrices $\vec\tau$ to denote gauge rotations for the Nambu index, in contrast to $\vec\sigma$ for spin rotations. It's straightforward to show that the above symmetry implementations (\ref{sym:abk general}) are exactly the same as (\ref{sym:maj:T1})-(\ref{sym:maj:c6}) for Majorana fermions.

\subsection{Classification of zero-flux $U(1)$ spin liquids}

In this section, we classify all symmetric $U(1)$ spin liquids on the honeycomb lattice, which preserves symmetry group (\ref{sym group}) generated by two translations $T_{1,2}$, hexagon-centered rotation $C_6$ and magnetic mirror $\tilde M_h=M_{[1\bar10]}\cdot\bst$. The idea is to identify all possible gauge rotations $\{G_U(i)\}$ associated with symmetry operations $\{U\in SG\}$, up to the following gauge redundancy:
\bea\label{gauge redundancy}
\Psi_i\rightarrow\Psi_iW_i,~~~G_U\big(U(i)\big)\rightarrow W^\dagger_{U(i)}G_U(U(i))W^U_i,\\
\notag W_i^U\equiv U W_i U^{-1},~~~W_i\in SU(2).~~~
\eea

In a $U(1)$ spin liquid, the mean-field ansatz of spinons has the following form
\bea
\hat H_{MF}^{U(1)}=\sum_{i,j}f^\dagger_{i\alpha}u_{i\alpha,j\beta}f_{j\beta}.
\eea
In this so-called canonical gauge\cite{Wen2002}, there is an emergent global $U(1)$ gauge symmetry
\bea
f_{i\alpha}\rightarrow e^{\imth\theta}f_{i\alpha}\Longleftrightarrow\Psi_i\rightarrow\Psi_ie^{\imth\theta\tau_z},~~~0\leq\theta<2\pi.~~~
\eea
This group of gauge rotations that preserve the mean-field ansatz is called the ``invariant gauge group'' (IGG). The projective symmetry group which characterizes the symmetry implementations on fractionalized spinons in a spin lquid, is an extension (2nd group cohomology $\mathcal{H}^2$) of the symmetry group $G_s$ by the IGG:
\bea
G_s=PSG/IGG,~~~PSG\in\mathcal{H}^2(G_s,IGG).
\eea

Similar to the algebraic relations (\ref{psg:T1,T2})-(\ref{psg:c6,m}) for Kitaev $Z_2$ spin liquids, the gauge rotations $\{G_U(i)\}$ in a symmetric $U(1)$ spin liquid satisfy the following algebra:
\bea
\label{psg:u1:T1,T2}&G_{T_1}(x_1+1,x_2,s)G_{T_2}(x_1,x_2,s)G_{T_1}^{-1}(x_1+1,x_2-1,s)G_{T_2}^{-1}(x_1+1,x_2,s)=e^{\imth\phi_{12}\tau_z},\\
\label{psg:u1:m,T1}&G_{\tilde M_h}^{-1}(x_1+1,x_2,s)G_{T_1}^\ast(x_1+1,x_2,s)G_{\tilde M_h}(x_1,x_2,s)\big[G_{T_2}^{-1}(x_2,x_1+1,s)\big]^\ast=e^{\imth\phi_{m,1}\tau_z},\\
\label{psg:u1:m,T2}&G_{\tilde M_h^{-1}}(x_1,x_2+1,s)G_{T_2}(x_1,x_2+1,s)G_{\tilde M_h}(x_1,x_2,s)\big[G_{T_1}^{-1}(x_2+1,x_1,s)\big]^\ast=e^{\imth\phi_{m,2}\tau_z},\\
\notag&G_{C_6}^{-1}(x_1+1,x_2,s)G_{T_1}(x_1+1,x_2,s)G_{C_6}(x_1,x_2,s)G_{T_2}(x_1+x_2-s,1-x_1,1-s)\\
\label{psg:u1:c6,T1}&\cdot G_{T_1}^{-1}(x_1+x_2+1-s,-x_1,1-s)=e^{\imth\phi_{c,1}\tau_z},\\
\label{psg:u1:c6,T2}&G_{C_6}^{-1}(x_1,x_2+1,s)G_{T_2}(x_1,x_2+1,s)G_{C_6}(x_1,x_2,s)G_{T_1}^{-1}(x_1+x_2+1-s,1-x_1,1-s)=e^{\imth\phi_{c,2}\tau_z},~~\\
\label{psg:u1:m}&G_{\tilde M_h}(x_2,x_1,s)G^\ast_{\tilde M_h}(x_1,x_2,s)=e^{\imth\phi_m\tau_z},\\
\label{psg:u1:c6}&G_{C_6}(C_6^5(i))G_{C_6}(C_6^4(i))G_{C_6}(C_6^3(i))G_{C_6}(C_6^2(i))G_{C_6}(C_6(i))G_{C_6}(i)=e^{\imth\phi_c\tau_z},\\
&G_{C_6}(1-x_2,x_1+x_2-s,1-s)G_{\tilde M_h}(x_1,x_2,s)\big[G_{C_6}(x_2,x_1,s)G_{\tilde M_h}(x_1+x_2-s,1-x_2,1-s)\big]^\ast=e^{\imth\phi_{c,m}\tau_z}.\label{psg:u1:c6,m}
\eea
where all the $\phi$'s are $U(1)$-valued variables.

To identify $U(1)$ spin liquids in proximity to the Kitaev $Z_2$ spin liquid with zero flux per hexagon, we focus on the solutions with $\phi_{12}=0$. Making use of gauge transformation (\ref{gauge redundancy}) with $W_i=e^{\imth\theta_i\tau_z}$, one can always choose a proper gauge so that the solutions have the following form
\bea
&G_{T_1}(i)=G_{T_2(i)}\equiv1,~~~\forall~i;\\
&G_{\tilde M_h}(x_1,x_2,s)=\big[e^{\imth s\alpha_m}\imth\tau_x\big]^{n_m},~~\alpha_m=0,\pi;\\
&G_{C_6}(x_1,x_2,s)=(\imth\tau_x)^{n_c}e^{\imth\alpha_c(s)\tau_z}.
\eea
where $n_m,n_c=0,1$. Most of the $U(1)$-valued phase factors in (\ref{psg:u1:T1,T2})-(\ref{psg:u1:c6,m}) become zero by proper gauge fixing:
\bea
\phi_{m,1}=\phi_{m,2}=\phi_{c,1}=\phi_{c,2}=\phi_m=0.
\eea

All 14 gauge-inequivalent solutions can be categorized into 4 types of symmetric $U(1)$ spin liquids:

(1) 2 distinct $U1A$ states with $n_m=n_c=0$. By proper gauge choice they satisfy $\alpha_m=0$ and
\bea
&\notag\phi_c=\alpha_{c}(0)+\alpha_c(1)=0,\\
&\phi_{c,m}=\alpha_c(0)-\alpha_c(1)=k\pi,~~~k=0,1.
\eea
These lead to 2 different $U1A$ states with
\bea
G_{\tilde M_h}(i)\equiv1,~~G_{C_6}(x_1,x_2,s)=e^{\imth\frac{k\pi}2(1-2s)\tau_z}.~~~
\eea

(2) 6 distinct $U1B$ states with $n_m=0,~n_c=1$. After proper gauge choice, they satisfy $\alpha_m=0$ and
\bea
&\notag\phi_{c,m}=-\alpha_c(0)-\alpha_c(1)=0,\\
&3[\alpha_c(0)-\alpha_c(1)]+\pi=\phi_c=0,\pi.
\eea
which lead to
\bea
\alpha_c(0)=-\alpha_c(1)=\frac{k\pi}{6},~k=0,1,2,3,4,5.
\eea
Therefore the gauge rotations for these $U1B$ states write
\bea
G_{\tilde M_h}(i)\equiv1,~~~G_{C_6}(x_1,x_2,s)=\imth\tau_x\cdot e^{\imth\frac{k\pi}{6}(1-2s)\tau_z}.~~~
\eea

(3) 2 distinct $U1C$ states with $n_m=1,~n_c=0$. After proper gauge choice, they satisfy
\bea
&\notag\phi_m=\phi_c=\alpha_c(0)=\alpha_c(1)=0,\\
&\alpha_m=\phi_{c,m}=k\pi,~~~k=0,1.
\eea
Therefore the gauge rotations for these $U1C$ states write
\bea
G_{\tilde M_h}(x_1,x_2,s)=(-1)^{ks}\imth\tau_x,~~~G_{C_6}(i)\equiv1.~~~
\eea

(4) 4 distinct $U1D$ states with $n_m=n_c=1$. After proper gauge choice, they satisfy
\bea
&\notag\alpha_m=k_1\pi,~~~k_1=0,1\\
&\notag\phi_{c,m}+\alpha_m=\alpha_c(1-s)-\alpha_c(s)=k_2\pi,~~k_2=0,1,\\
&3[\alpha_c(0)-\alpha_c(1)]+\pi=\phi_c=(k_2+1)\pi.
\eea
which lead to (after gauge fixing)
\bea
\alpha_c(0)=0,~~\alpha_c(1)=k_2\pi.
\eea
Therefore the gauge rotations for these $U1B$ states write
\bea
G_{\tilde M_h}(x_1,x_2,s)=(-1)^{k_1s}\imth\tau_x,\\
G_{C_6}(x_1,x_2,s)=(-1)^{k_2s}\imth\tau_x.
\eea
In total there are 2+6+2+4=14 distinct $U(1)$ spin liquids that preserve symmetry group (\ref{sym group}).

In the anisotropic model, $C_6$ symmetry is broken while preserving the inversion symmetry $I=(C_6)^3$ and the symmetry group becomes
\bea\label{sym group:anisotropy}
G_s=\{T_1^{\nu_1}T_2^{\nu_2}\tilde M_h^{\nu_m}I^{\nu_I}|\nu_{1,2}\in\mbz,\nu_m,\nu_I\in\mbz_2\}.~~~~
\eea
It's straightforward to show that the 6 distinct $U1B$ states preserving $C_6$ symmetry now collapse into only 2 distinct $U1B$ states with inversion symmetry $I$. More precisely, with only inversion symmetry $I$, the $k=0\mod2$ solutions become one $U1B$ state while $k=1\mod2$ solutions become the other. All other states remain distinct when breaking $C_6$ symmetry down to inversion $I$. This leads to 2+2+2+4=10 distinct symmetric $U(1)$ spin liquids for the anisotropic case.

\subsection{$U(1)$ spin liquids neighboring the Kitaev $Z_2$ states}

Once the $IGG=U(1)$ gauge group is broken down a $Z_2$ subgroup by a pairing term between fermionic spinons $\{f_{i\uparrow},f_{i\downarrow}\}$, the $U(1)$ spin liquid is driven into a $Z_2$ spin liquid via a Higgs transition. Among all possible $U(1)$ spin liquids, which ones are related to the Kitaev $Z_2$ spin liquid by a continuous Higgs transition?

The gauge rotations on fermionic spinons in the Kitaev $Z_2$ spin liquids satisfy algebra (\ref{psg:T1,T2})-(\ref{psg:c6,m}). For any $U(1)$ spin liquid connected with Kitaev $Z_2$ states by a continuous quantum phase transition, the spinon PSGs must be compatible with the $Z_2$ state. Specifically in the canonical gauge, the $U(1)$ PSGs can always be redefined by a global $U(1)$ gauge rotation
\bea
G_U(i)\rightarrow e^{\imth\gamma_U\tau_z}G_U(i)
\eea
Meanwhile $\{G_U(i)\}$ also has a gauge redundancy shown in (\ref{gauge redundancy}). When gauge rotations associated with both translations are fixed as
\bea
G_{T_{1,2}}(i)\equiv1,~~~\forall~i.~~~
\eea
the only remaining gauge redundancies are the sublattice gauge rotations $W(x_1,x_2,s)=W_s$ in (\ref{gauge redundancy}). Therefore for any $U(1)$ spin liquid proximate to the Kitaev $Z_2$ spin liquid with gauge transformations (\ref{sym:abk:T1,2})-(\ref{sym:abk:c6}), its gauge transformations $\{G_U(i)\}$ must be related to (\ref{sym:abk:T1,2})-(\ref{sym:abk:c6}) by a gauge choice:
\bea\label{z2 to u1:mirror}
&(-1)^s W_s^\dagger e^{-\imth\frac\pi4\tau_z}W_s^\ast=e^{\imth\gamma_m\tau_z}G_{\tilde M_h}(x_1,x_2,s),~~~\\
\label{z2 to u1:rot}&(-1)^s W_s^\dagger e^{\imth\frac{\pi}{3}\frac{\tau_x+\tau_y+\tau_z}{\sqrt3}}W_{1-s}=e^{\imth\gamma_c\tau_z}G_{C_6}(x_1,x_2,s).~~~~~
\eea
These two equations can be rewritten as
\bea
&W_s^\dagger\frac{\tau_y-\tau_x}{\sqrt2}W_s=(-1)^{s}e^{\imth\gamma_m\tau_z}G_{\tilde M_h}(x_1,x_2,s)\tau_y,~~~\\
&W_0^\dagger e^{\imth\frac{\pi}{3}\frac{\tau_x+\tau_y+\tau_z}{\sqrt3}}W_{1}=e^{\imth\gamma_c\tau_z}G_{C_6}(x_1,x_2,0),~~~\\
\notag&W_0^\dagger e^{-\imth\frac{\pi}{3}\frac{\tau_x+\tau_y+\tau_z}{\sqrt3}}W_0=\\
&e^{\imth\gamma_c\tau_z}G_{C_6}(x_1,x_2,0)e^{\imth\gamma_c\tau_z}G_{C_6}(x_1,x_2,1).
\eea

It is straightforward to show that no $U1C$ or $U1D$ states satisfy the above conditions. There are only 3 symmetric $U(1)$ spin liquids in proximity to the Kitaev $Z_2$ states with (\ref{sym:abk:T1,2})-(\ref{sym:abk:c6}), as summarized below:

(1) $U1A_{k=0}$ state with
\bea\label{U1A}
G_{\tilde M_h}(i)=G_{C_6}(i)=G_I(i)\equiv1,~~~\forall~i.
\eea
The solution of (\ref{z2 to u1:mirror})-(\ref{z2 to u1:rot}) is given by
\bea
&W_0=U_0\cdot(\imth\tau_z),~~~W_1=U_0,\\
&U_0\equiv e^{\frac\imth2(\arccos{\frac1{\sqrt3}})\frac{\tau_x-\tau_y}{\sqrt2}}.\label{z2 to u1: gauge rot}
\eea
and
\bea
\gamma_m=\frac34\pi,~~~\gamma_c=-\frac16\pi.~~
\eea

(2) $U1B_{k=2}$ state with
\bea
&\notag G_{\tilde M_h}(i)=1,~~G_{C_6}(x_1,x_2,s)=\imth\tau_x\cdot e^{\imth\frac{\pi}{3}(1-2s)\tau_z},\\
&G_I(i)\equiv\imth\tau_x,~~~\forall~i.~~~\label{U1B2}
\eea
The solution of (\ref{z2 to u1:mirror})-(\ref{z2 to u1:rot}) is given by
\bea
&W_0=U_0\cdot e^{\imth\frac{3\pi}4\tau_z}(\imth\tau_x),~~~W_1=U_0\cdot e^{\imth\frac{3\pi}4\tau_z},\\
&U_0\equiv e^{\frac\imth2(\arccos{\frac1{\sqrt3}})\frac{\tau_x-\tau_y}{\sqrt2}}.
\eea
and
\bea
\gamma_m=0,~~~\gamma_c=\pi.~~
\eea

(3) $U1B_{k=4}$ state with
\bea
&\notag G_{\tilde M_h}(i)=1,~~G_{C_6}(x_1,x_2,s)=\imth\tau_x\cdot e^{-\imth\frac{\pi}{3}(1-2s)\tau_z},\\
&G_I(i)\equiv\imth\tau_x,~~~\forall~i.~~~\label{U1B4}
\eea
The solution of (\ref{z2 to u1:mirror})-(\ref{z2 to u1:rot}) is given by
\bea
&W_0=U_0\cdot e^{-\imth\frac{\pi}4\tau_z},~~~W_1=U_0\cdot e^{-\imth\frac{\pi}4\tau_z}(\imth\tau_x),\\
&U_0\equiv e^{\frac\imth2(\arccos{\frac1{\sqrt3}})\frac{\tau_x-\tau_y}{\sqrt2}}.
\eea
and
\bea
\gamma_m=0,~~~\gamma_c=\pi.~~
\eea

Note that when $C_6$ rotational symmetry is broken down to inversion by anisotropy $J_z\neq J_{x,y}$, the two states $U1B_{k=2}$ and $U1B_{k=4}$ collapse into the same $U(1)$ spin liquid phase.

\subsection{Stability of spinon fermi surfaces}

Below we use K-theory classification\cite{Kitaev2009,Teo2010,Matsuura2013,Chiu2016} to analyze the stability of spinon fermi surfaces and Dirac points in the three $U(1)$ states in proximity to Kitaev $Z_2$ states. Since the gapless phase persists even in the presence of $C_6$-breaking anisotropy in the phase diagram, the symmetries considered here will only include inversion $I$, magnetic mirror $\tilde M_h=\bst\cdot M_{[1\bar10]}$ and two translations $T_{1,2}$.

(1) $U1A_{k=0}$ state. According to (\ref{U1A}), under symmetry operations the spinons transform in the same way as the electrons in a spin-orbit coupled magnetic metal. The stable fermi surfaces is classified by $\pi_0(\mathcal{C}_0)=\mbz$, where $\mathcal{C}_0$ is the classifying space for zero-dimensional (at each ${\bf k}$) gapped Hamiltonians. The integer topological index $\nu\in\mbz$ labels the change of spinon filling (below fermi energy) at each ${\bf k}$ across the fermi surface. Meanwhile, at a generic momentum with no extra symmetries in the 1st BZ, the stable Dirac points are classified by $\pi_1(\mathcal{C}_0)=0$. Meanwhile since $(\tilde M_h)^2=(I\cdot\tilde M_h)^2=1$, the stable Dirac points on high symmetry lines $k_{x,y}=0,\pi$ are classified by $\pi_0(\mathcal{R}_{2-3})=0$.

As a result, $U1A_{k=0}$ state supports robust spinon fermi surfaces labeled by an integer index $\nu\in\mbz$, but not Dirac points of spinons.\\

(2) $U1B_{k=2}$ state. According to (\ref{U1B2}), the gauge rotation $G_I(i)=\imth\tau_x$ associated with inversion symmetry $I$ is a particle-hole transformation, which anticommutes with the generator $\imth\tau_z$ of the global $U(1)$ IGG:
\bea
\{G_I(i),\imth\tau_z\}=0.
\eea
Since inversion also reverses momentum ${\bf k}$, under the inversion operation the spinons transform as
\bea
\bpm f_{{\bf k},\uparrow}\\f_{{\bf k},\downarrow}\epm\overset{I}\longrightarrow\imth\bpm f^\dagger_{{\bf k},\downarrow}\\-f^\dagger_{{\bf k},\uparrow}\epm
\eea
This ``particle-hole symmetry'' at every single ${\bf k}$ leads to a classifying space of $\mathcal{R}_{0-2-d}=\mathcal{R}_{6-d}$ for a gapped Hamiltonian in $d$ spatial dimensions. Therefore stable spinon fermi surfaces for $U1B_{k=2}$ state are classified by $\pi_0(\mathcal{R}_6)=0$. Similarly stable Dirac points at a generic momentum are classified by $\pi_0(\mathcal{R}_5)=0$. Meanwhile, since $[\tilde M_h,I]=0$ in $U1B_{k=2}$ state, Dirac points located on high symmetry lines $k_y=0,\pi$ (invariant under $\tilde M_h$) are classified by $\pi_0(\mathcal{R}_{2-4})=0$.

As a result, $U1B_{k=2}$ state supports neither stable fermi surfaces or Dirac points.\\

(3) $U1B_{k=4}$ state. This state is completely similar to $U1B_{k=2}$ state discussed previously, and hosts neither stable fermi surfaces or Dirac points of fermionic spinons.

\begin{figure}
		[h]
		\parbox{4.2cm}{
		\boxed{
		\includegraphics[width=4cm]{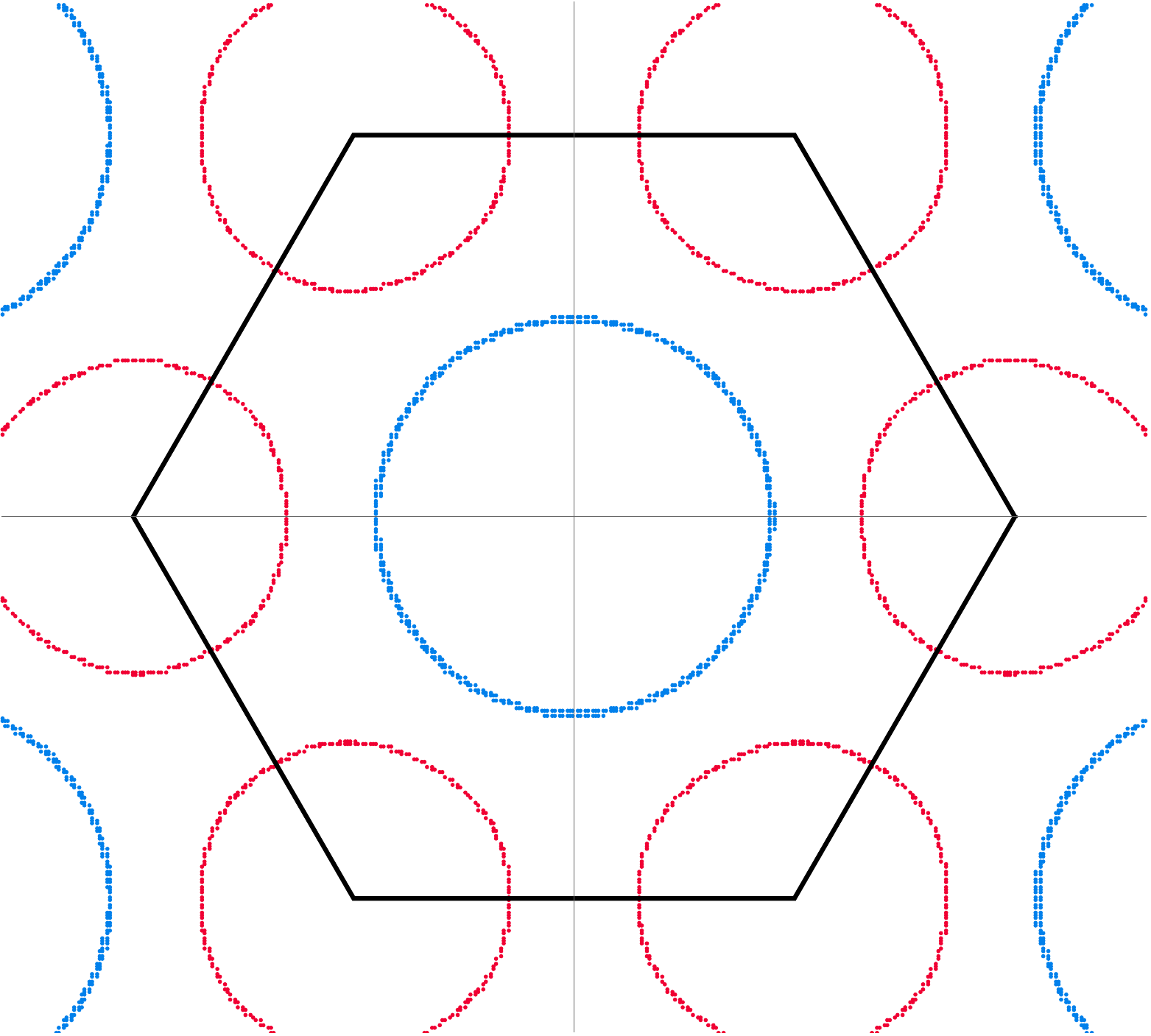}}\\
	(a) $h=0.25,~J_z/J=1$
	}
\parbox{4.2cm}{
		\boxed{
		\includegraphics[width=4cm]{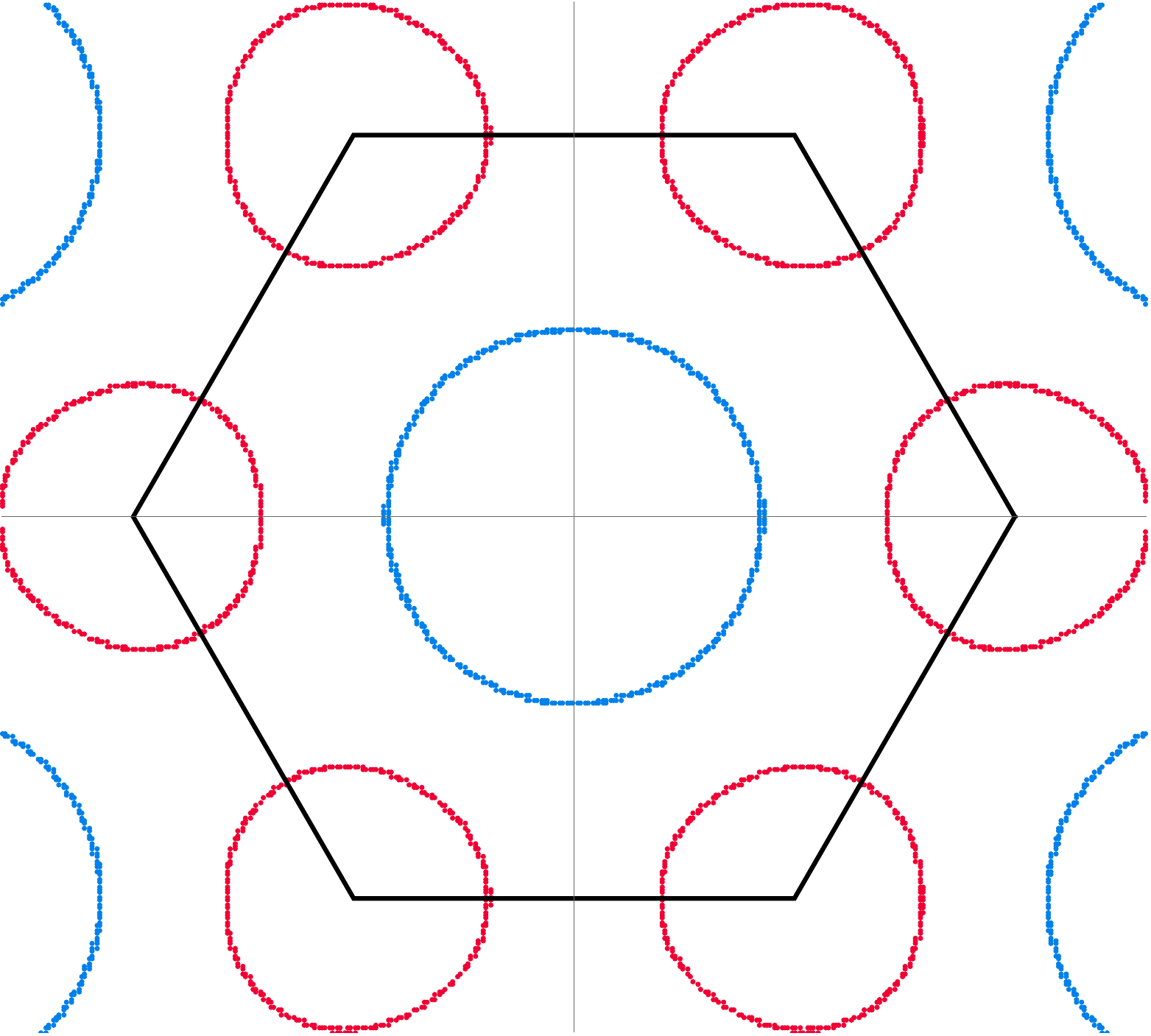}}\\
	(b) $h=0.3,~J_z/J=1$
    }
	\\ \qquad \\ \qquad \\
\parbox{4.2cm}{
		\boxed{\includegraphics[width=4cm]{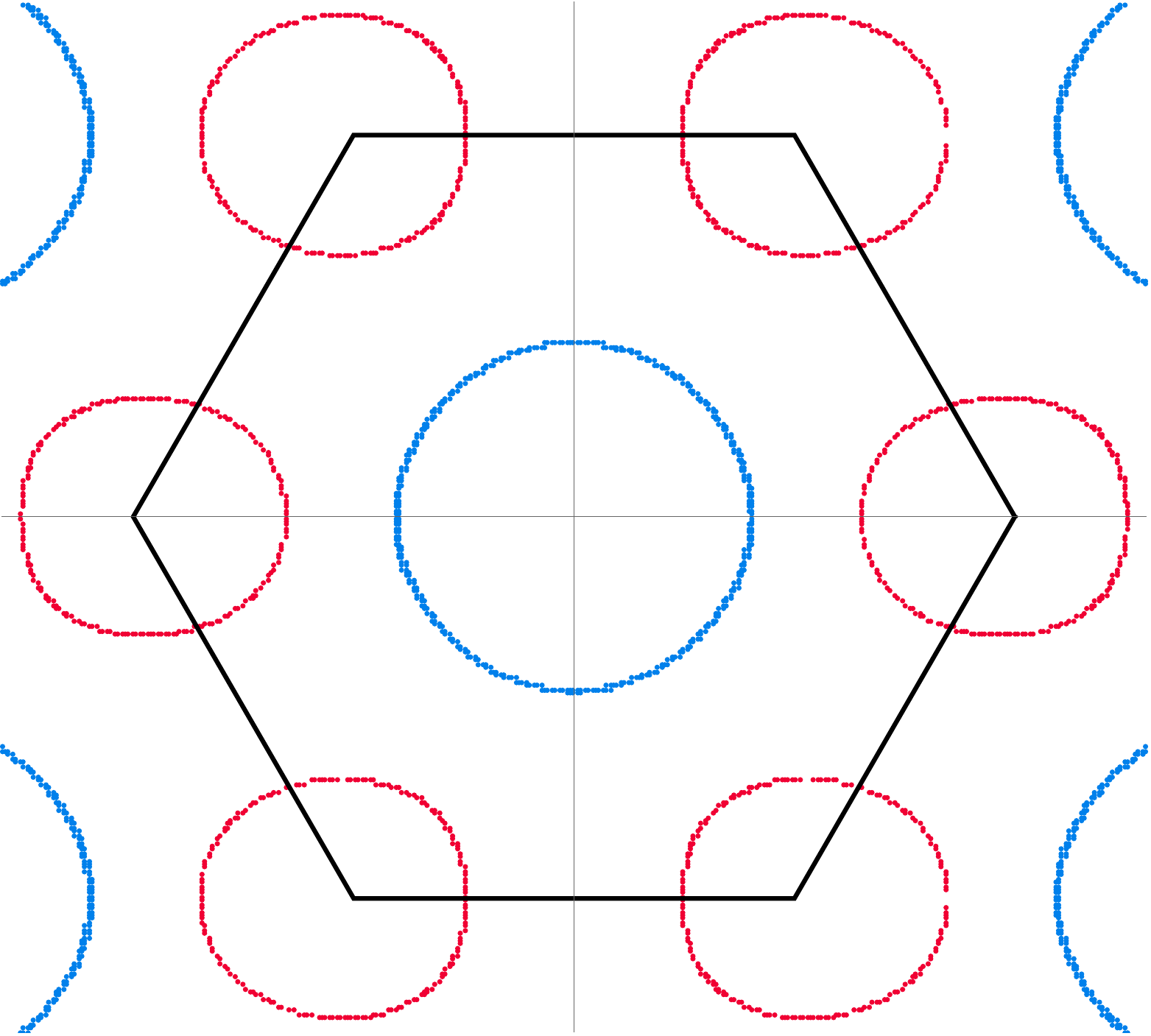}}\\
    (c) $h=0.3,~J_z/J=2$
	}
\parbox{4.2cm}{
		\boxed{\includegraphics[width=4cm]{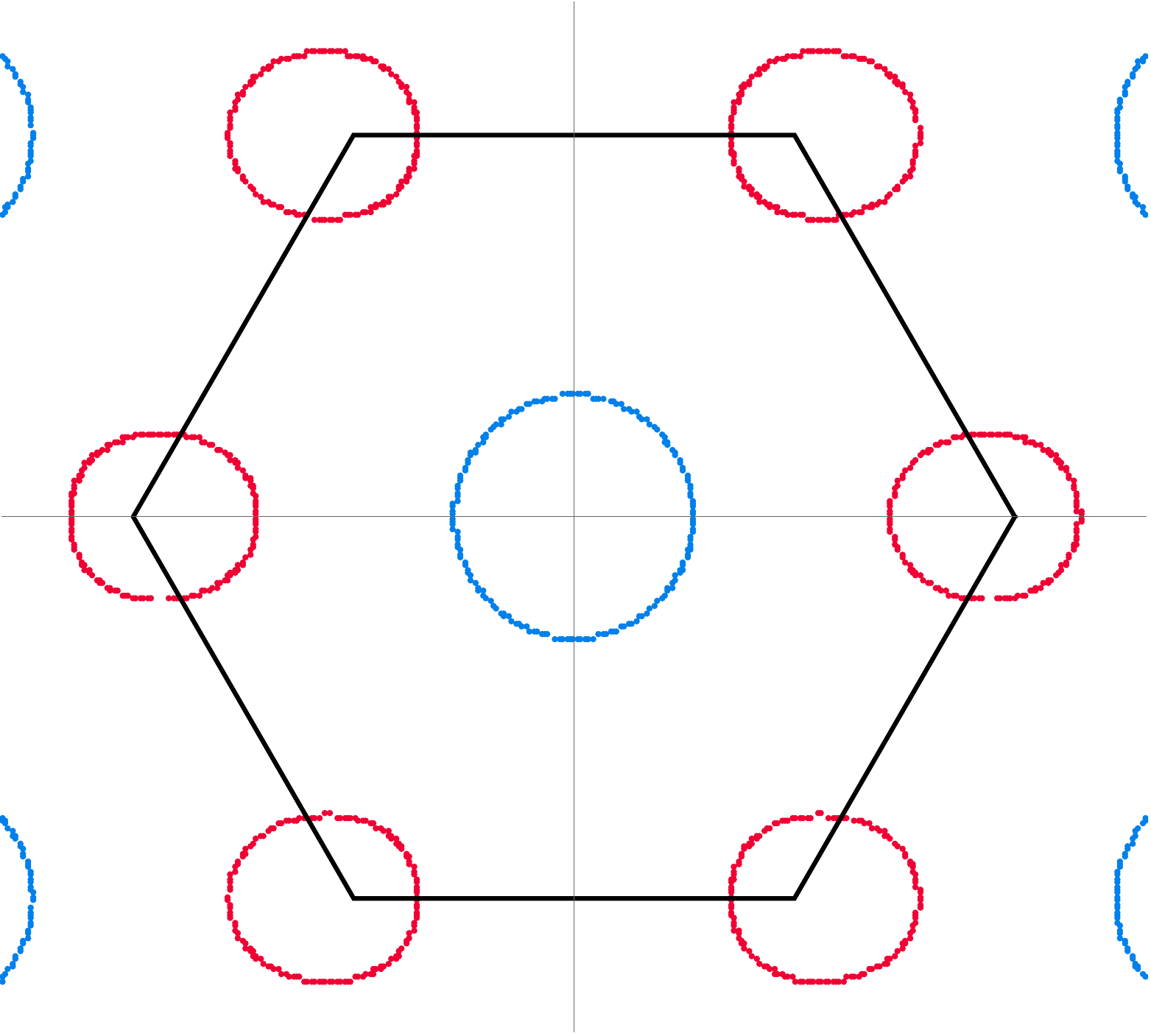}}\\ (d) $h=0.3,~J_z/J=3$
	}
\caption{(Color online) Evolution of spinon fermi surfaces (FS) with increased [111] magnetic field $h$ and anisotropy $J_z/J$. The blue circle denotes one electron pocket around zone center $\Gamma$, and red circles denote two hole pockets around zone corners $K$ and $K^\prime$, where the 1st Brillouin zone (BZ) is a hexagon colored in black. The non-zero mean field parameters we use are $ (s_3, t^x_0, t^y_0) = (-0.019, -0.19, -0.0079)  $ for $\alpha, \beta$ bonds, $ (J_z/J)(s_3, t^x_0, t^y_0) $ for $\gamma$ bonds, and $ (\tilde{s}_0, \tilde{s}_3, \tilde{t}^x_3, \tilde{t}^z_3) = (-0.0055, 0.0018, -0.0032, -0.0032) $ for NNN bonds. The chemical potential is tuned such that the system is at half-filling. The magnetic field $h$ and anisotropy $J_z/J$ are specified in each subplots.} \label{fig:spinon fs:app}
	\end{figure}

\begin{figure}
\includegraphics[width=8cm]{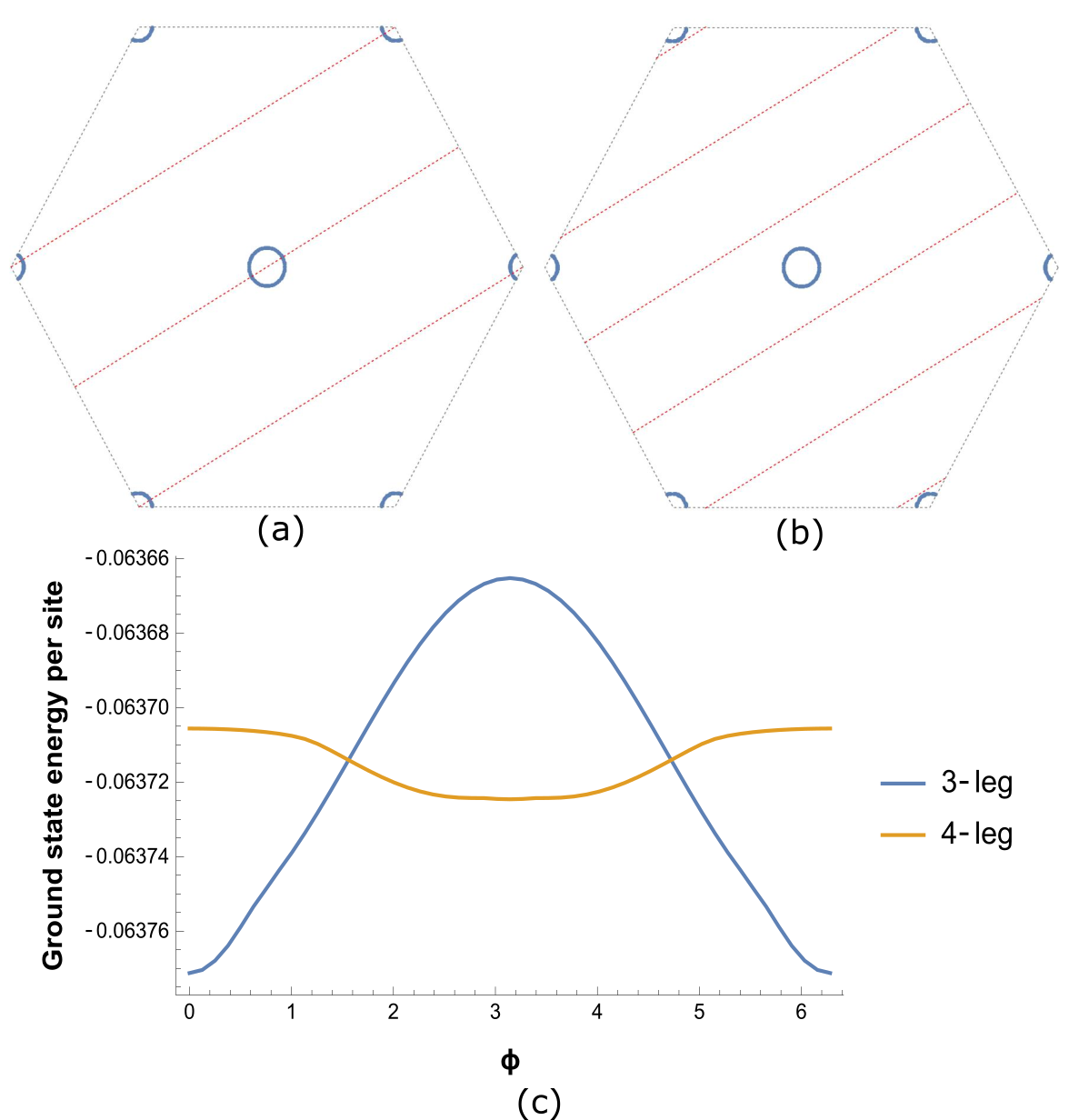}
\caption{The intersections between spinon fermi surfaces (FSs) and quantized momentum around the cylinder on the 3-leg ladder (a) versus 4-leg ladder (b). The parameters in the mean-field ansatz are chosen as $(s_0, t_0^x, t_0^y)=(-0.02373, -0.02373, -0.00415),\  (\tilde{s}_0, \tilde{s}_3, \tilde{t}^x_3, \tilde{t}^z_3)=(-0.00018, 0.0018, -0.000102,  -0.000102)$, with a magnetic field $h=0.188$ and a chemical potential $\mu=-0.01832$. (c) The dependence of spinon mean-field energy as a function of boundary condition around the cylinder. On the 3-leg ladder, periodic boundary condition has the lowest energy, where ground state energy per site is $-0.06377$. On the 4-leg ladder, anti-periodic boundary condition has the lowest energy, where ground state energy per site is $-0.06373$.}\label{fig:spinon fs:ladders}
\end{figure}

\section{Mean-field ansatz of the U1A state}\label{app:U1A state mfh}

We follow the Fourier transformation convention
	\begin{align}
	f_{i\alpha} = \frac{1}{\sqrt{N_{\text{cell}}}} \sum_{\boldsymbol{k}} e^{-i\boldsymbol{k}\cdot \boldsymbol{R}_i} f_{\boldsymbol{k}\alpha},
	\end{align}
	where $ \boldsymbol{R}_i $ is the unit-cell position, and the {\bf k}-space	basis
	\begin{align}
	\Psi_{\boldsymbol{k}} = (a_{\boldsymbol{k}\uparrow}, a_{\boldsymbol{k}\downarrow}, b_{\boldsymbol{k}\uparrow}, b_{\boldsymbol{k}\downarrow})^T,
	\end{align}
($a_{\bf k}$ and $b_{\bf k}$ for A/B sublattices respectively) the spinon mean-field Hamiltonian has the following form
	\begin{align}
	H &= \sum_{\boldsymbol{k}} \Psi_{\boldsymbol{k}}^\dagger h_{\boldsymbol{k}} \Psi_{\boldsymbol{k}},\\
	h_{\boldsymbol{k}} &= h_{0\boldsymbol{k}} + h_{1\boldsymbol{k}} + h_{2\boldsymbol{k}},
	\end{align}
	where 0,1,2 denote onsite, N.N., and N.N.N. terms, as defined below.

Labeling 2d momentum by ${\bf k}=k_1\vec b_1+k_2\vec b_2$ where $\vec b_{1,2}$ are reciprocal lattice vectors associated with Bravais lattice vectors $\vec a_{1,2}$ we have
\begin{align}
	k_1 = \frac{\sqrt3 k_x + 3k_y}{2},\qquad
	k_2 = \frac{-\sqrt3 k_x + 3k_y}{2}.
	\end{align}

The onsite terms are given by
	\begin{align}
	h_{0} = -\mu \tau_0\sigma_0 - \frac{h}{8} \tau_0 (\sigma_x+\sigma_y + \sigma_z)
	\end{align}

NN terms are
	\begin{align}
	h_{1\boldsymbol{k}} &= - \begin{pmatrix}
	0 & D_{\boldsymbol{k}} \\ D_{\boldsymbol{k}}^\dagger & 0
	\end{pmatrix}\\ \nonumber
	D_{\boldsymbol{k}} &= (s_3\sigma_0 + t^x_0\sigma_x + t^y_0\sigma_y + t^y_0\sigma_z) e^{-ik_1}\\ \nonumber
	&  + (s_3\sigma_0 + t^y_0\sigma_x + t^x_0\sigma_y + t^y_0\sigma_z) e^{-ik_2}\\
	&  + (s_3\sigma_0 + t^y_0\sigma_x + t^y_0\sigma_y + t^x_0\sigma_z)
	\end{align}

and NNN terms are
	\begin{align}
	h_{2\boldsymbol{k}} =&\, -\begin{pmatrix}
	A_{\boldsymbol{k}} & 0 \\ 0 & B_{\boldsymbol{k}}
	\end{pmatrix},\\ \nonumber
	A_{\boldsymbol{k}} =&\, 2 (\tilde{s}_0\sigma_0 + \tilde{t}^x_3\sigma_x + \tilde{t}^x_3 \sigma_y + \tilde{t}^z_3\sigma_z) \sin(k_1-k_2)\\ \nonumber
	& + 2(\tilde{s}_3\sigma_0 + \tilde{t}^x_0\sigma_x + \tilde{t}^x_0\sigma_y + \tilde{t}^z_0\sigma_z) \cos(k_1-k_2)\\
	\nonumber
	& + 2 (\tilde{s}_0 \sigma_0 + \tilde{t}^z_3 \sigma_x + \tilde{t}^x_3 \sigma_y + \tilde{t}^x_3 \sigma_z) \sin (-k_2)\\ \nonumber
	& + 2 (\tilde{s}_3 \sigma_0 + \tilde{t}^z_0\sigma_x + \tilde{t}^x_0 \sigma_y + \tilde{t}^x_0\sigma_z) \cos (-k_2)\\ \nonumber
	& + 2 (\tilde{s}_0\sigma_0 + \tilde{t}^x_3 \sigma_x + \tilde{t}^z_3 \sigma_y + \tilde{t}^x_3\sigma_z) \sin (k_1)\\
	& + 2 (\tilde{s}_3+ \tilde{t}^x_0\sigma_x + \tilde{t}^z_0 \sigma_y + \tilde{t}^x_0 \sigma_3) \cos (k_1),\\ \nonumber
	B_{\boldsymbol{k}} =&\, 2(\tilde{s}_0\sigma_0 + \tilde{t}^x_3\sigma_x + \tilde{t}^z_3\sigma_y + \tilde{t}^x_3\sigma_z) \sin(-k_1)\\ \nonumber
	& + 2(\tilde{s}_3\sigma_0 + \tilde{t}^x_0\sigma_x + \tilde{t}^z_0\sigma_y + \tilde{t}^x_0\sigma_z) \cos(-k_1)\\ \nonumber
	& + 2(\tilde{s}_0\sigma_0 + \tilde{t}^x_3\sigma_x + \tilde{t}^x_3\sigma_y + \tilde{t}^z_3\sigma_z) \sin[-(k_1-k_2)]\\ \nonumber
	& + 2(\tilde{s}_3 \sigma_0 + \tilde{t}^x_0\sigma_x + \tilde{t}^x_0 \sigma_y + \tilde{t}^z_0\sigma_z) \cos[-(k_1-k_2)]\\ \nonumber
	& + 2(\tilde{s}_0 \sigma_0 + \tilde{t}^z_3\sigma_x + \tilde{t}^x_3\sigma_y + \tilde{t}^x_3\sigma_z) \sin(k_2) \\
	& + 2(\tilde{s}_3 \sigma_0 + \tilde{t}^z_0\sigma_x + \tilde{t}^x_0 \sigma_y + \tilde{t}^x_0\sigma_z) \cos(k_2).
	\end{align}

In FIG. \ref{fig:spinon fs:app} we demonstrate the shrinking of spinon fermi surfaces as we increase bond anisotropy $J_z/J$ and magnetic field $h_{[111]}$.

In FIG. \ref{fig:spinon fs:ladders} we show in the isotropic case with $J_z/J=1$, how the spinon fermi surfaces intersect with quantized momenta along the circumference of the cylinder, on 3-leg and 4-leg ladders. On a 3-leg ladder, periodic boundary condition of spinons along the circumference minimizes the energy, where the spinon FSs intersect with the quantized momenta at both $\Gamma$ and $\pm K$. This leads to nonzero central charge $c\leq2$ on a 3-leg ladder. On a 4-leg ladder, however, anti-periodic boundary condition minimizes the spinon energy, leading to no crossing between spinon FSs with quantized momenta, and hence a zero central charge on the 4-leg ladder.

\section{Topological invariant of 2d centrosymmetric superconductors in class D}\label{app:topo inv}

Here we discuss the topological invariant of a gapped superconductor in two spatial dimensions. We consider symmetry class D, \ie superconductors with neither time reversal nor spin rotational symmetries.

A generic BdG Hamiltonian for a class-D superconductor is written as
\bea
\hat H_{\text{BdG}}=\frac12\sum_{\bf k}\psi^\dagger_{\bf k}H_{\bf k}\psi_{\bf k},~~H_{\bf k}=\bpm h_{\bf k}&\Delta_{\bf k}\\ \Delta^\dagger_{\bf k}&-h^\ast_{-{\bf k}}\epm.\label{BdG:class D}
\eea
where we defined spinor
\bea
\psi_{\bf k}=(f_{{\bf k},\uparrow},f_{{\bf k},\downarrow},f^\dagger_{-{\bf k},\uparrow},f^\dagger_{-{\bf k},\downarrow})^T=\tau_x\psi^\ast_{-{\bf k}}.
\eea
In two spatial dimensions (2d), the integer-valued topological invariant for a superconductor in symmetry class D is given by the Chern number $\nu\in\mbz$ of BdG Hamiltonian (\ref{BdG:class D}). It is also the number of chiral edge modes on the open boundary of the superconductor\cite{Read2000}.

In the presence of inversion symmetry $\hat I$ with $\hat I^2=(-1)^{\hat F}$, the parity of Chern number $\nu$ is determined by the inversion eigenvalues $\{P_j({\bf k})=\pm\imth\}$ of all filled bands at the 4 time reversal invariant momenta (TRIM)\cite{Turner2010,Hughes2011,Fang2012}:
\bea\label{parity invariant}
e^{\imth\pi\nu}=(-1)^\nu=\prod_{E_{{\bf k},j}<0}\Big[\prod_{{\bf k}\in\text{TRIM}}P_j({\bf k})\Big],\\
\notag\hat I\dket{{\bf k},j}=P_j({\bf k})\dket{{\bf k},j},~~~{\bf k}\in\text{TRIM}.
\eea

Now let's consider the onset of a small pairing $\Delta_{\bf k}$ on top of a fermion band structure described by a Bloch Hamiltonian $h_{\bf k}$, where $|\Delta_{\bf k}|\ll|h_{\bf k}|$ for each ${\bf k}\in$~TRIM.
%As discussed earlier, $U1A_{k=0}$ state in TABLE \ref{tab:psg} is the only candidate for the gapless $U(1)$ QSL phase in Kitaev model under [111] field. In $U1A_{k=0}$ the inversion symmetry $\hat I$ is implemented by
Without loss of generality we can always choose the following implementation of inversion symmetry $\hat I$:
\bea
\hat I f_{\bf k}\hat I^{-1}=\imth U_I f_{-\bf k},~~~(U_I)^\ast=U_I=(U_I)^{-1}=(U_I)^T.
\eea
and hence
\bea
\hat I\psi_{k}\hat I^{-1}=(\imth U_I\otimes\tau_z)\psi_{-{\bf k}}
\eea
in the Nambu basis. As a result the inversion eigenvalues can be computed by
\bea
\imth U_I\otimes\tau_z{\bf V}_{{\bf k},j}=P_j({\bf k}){\bf V}_{{\bf k},j},~~~\forall~{\bf k}\in\text{TRIM}.
\eea
for any BdG eigenvector ${\bf V}_{\bf k}$ at the TRIM
\bea
H_{\bf k}{\bf V}_{{\bf k},j}=E_{{\bf k},j}{\bf V}_{{\bf k},j}.
\eea

Let's first consider the parity invariant (\ref{parity invariant}) for a trivial band insulator (with zero Chern number $C_0=0$) where $\Delta_{\bf k}\equiv0$. Assuming Bloch eigenvectors ${\bf v}_{{\bf k},j}$ with energy $\epsilon_{{\bf k},j}$
\bea
h_{\bf k}{\bf v}_{{\bf k},kj}=\epsilon_{{\bf k},j}{\bf v}_{{\bf k},j}
\eea
its parity eigenvalues of all filled bands at TRIM must multiply to be unity:
\bea
e^{\imth\pi C_0}=\prod_{\epsilon_{{\bf k},j}<0}\big[\prod_{{\bf k}\in\text{TRIM}}p_j({\bf k})\big]=1,\\
\imth U_I{\bf v}_{{\bf k},j}=p_j({\bf k}){\bf v}_{{\bf k},j},~~~\forall~{\bf k}\in\text{TRIM}.
\eea
Note that all filled bands and all unfilled bands must have opposite Chern numbers and therefore
\bea
e^{-\imth\pi C_0}=\prod_{\epsilon_{{\bf k},j}>0}\big[\prod_{{\bf k}\in\text{TRIM}}p_j({\bf k})\big]=1
\eea
The corresponding BdG eigenvectors for this trivial insulator are given by
\bea
{\bf V}_{{\bf k},2j}=\bpm{\bf v}_{{\bf k},2j}\\0\epm,~~E_{{\bf k},2j}=\epsilon_{{\bf k},j},\\
P_{2j}({\bf k})=p_j({\bf k})=\pm\imth\notag
\eea
and
\bea
{\bf V}_{{\bf k},2j+1}=\bpm0\\{\bf v}^\ast_{{\bf k},2j}\epm,~~E_{{\bf k},2j+1}=-\epsilon_{{\bf k},j},\\
\notag P_{2j+1}({\bf k})=p^\ast_j({\bf k})=-p_j({\bf k}).
\eea
Therefore the parity invariant (\ref{parity invariant}) for BdG Hamiltonian (\ref{BdG:class D}) is given by
\bea
e^{\imth\pi\nu}=e^{2\imth\pi C_0}=1
\eea
in the case of a trivial band insulator. Starting from this trivial band insulator, the appearance of a single electron/hole pocket at one TRIM corresponds to a single band inversion for the corresponding BdG Hamiltonian, which reverses the sign of the parity invariant (\ref{parity invariant}). As a result for an arbitrary Bloch Hamiltonian $h_{\bf k}$ with a finite energy gap at every TRIM, the parity invariant (\ref{parity invariant}) of the corresponding BdG Hamiltonian (\ref{BdG:class D}) is given by
\bea
e^{\imth\pi\nu}=(-1)^{\#~\text{of pockets at all TRIM points}}
\eea
or the Chern number $\nu$ of the BdG Hamiltonian is given by
\bea\label{parity inv:spinon fs}
\nu=\#~\text{of pockets at all TRIM points}\mod 2.
\eea
in the presence of inversion symmetry $\hat I$.

Now let's consider the onset of a small pairing order parameter $\Delta_{\bf k}$ on top of the band structure $h_{\bf k}$, such that
\bea
|\Delta_{\bf k}|\ll|\epsilon_{{\bf k},j}|,~~~\forall~j,~{\bf k}\in\text{TRIM}.
\eea
While this small pairing order parameter opens up a gap for the BdG Hamiltonian in the while BZ, it cannot change the parity eigenvector (\ref{parity inv:spinon fs}) of the BdG Hamiltonian. As a result, the Chern number $\nu$ of a gapped BdG Hamiltonian, in the small pairing limit, is determined by the total number of fermi surfaces around all TRIM.

\end{widetext}

\end{document}